# Energy Loss Distribution in the Plane Couette Flow and the Taylor-Couette Flow between Concentric Rotating Cylinders

Hua-Shu Dou[1,2,*], Boo Cheong Khoo[2], and Khoon Seng Yeo[2]

[1]Temasek Laboratories, National University of Singapore, Singapore 117508, SINGAPORE

[2]Department of Mechanical Engineering

National University of Singapore, Singapore 119260, SINGAPORE

*Corresponding Author, Email: tsldh@nus.edu.sg; huashudou@yahoo.com

**Abstract** The distribution of energy loss due to viscosity friction in plane Couette flow and Taylor-Couette Flow between concentric rotating cylinders are studied in detail for various flow conditions. The energy loss is related to the industrial processes in some fluid delivery devices and has significant influence on the flow efficiency, flow stability, turbulent transition, mixing, and heat transfer behaviours, etc. Therefore, it is important to know about the energy loss distribution in the flow domain and to know its influence on the flow for better understanding of the flow physics. The calculation or methodology of calculating the energy loss distribution in the Taylor-Couette Flow between concentric rotating cylinders are not readily found in the open literature. In this paper, the principle and the calculation are given for single cylinder rotation of either the inner or outer cylinder, and counter and same direction rotation of two cylinders. For comparison, the distribution of energy loss in a plane Couette flow is also derived for various flow conditions. Discussions of the effect of energy loss on the flow behaviour are carried out from which some findings are suggested.

## Nomenclature

| | | |
|---|---|---|
| $A_1$ | work done to the element by the upper layer in Fig.3 | J |
| $A_2$ | work done to the lower layer by the element in Fig.3 | J |

E total energy of unit volume of fluid $J\ m^{-3}$

g acceleration of gravity $m\ s^{-2}$

h width of channel for the plane Couette flow m

H energy loss of unit volume of fluid due to viscosity in streamwise direction $J\ m^{-3}$

$H_t$ energy consumed by the fluid in the system $J\ m^{-3}$

K kinetic energy in unit volume of fluid $J\ m^{-3}$

p static pressure $N\ m^{-2}$

P potential energy in unit volume of fluid $J\ m^{-3}$

Q heat added to the system by external $J\ m^{-3}$

Q fluid volume passing through *dy* depth in *dt* time in Fig. 3 $m^3$

$R_1$ radius of inner cylinder m

$R_2$ radius of outer cylinder m

Re Reynolds number (dimensionless).

s length in streamwise direction m

u velocity component in the main flow direction $m\ s^{-1}$

$U_1$ velocity of upper plate in plane Couette flow $m\ s^{-1}$

$U_2$ velocity of bottom plate in plane Couette flow $m\ s^{-1}$

v velocity component in the transverse direction $m\ s^{-1}$

V total velocity $m\ s^{-1}$

W work done by external force $J\ m^{-3}$

$W_t$ work done on the fluid in the system by all external forces $J\ m^{-3}$

x coordinate in the streamwise direction m

y coordinate in the transverse direction m

z coordinate in the spanwise direction m

$\eta$ radius ratio, $\equiv R_2 / R_1$

$\theta$ angular coordinates rad

$\lambda$ speed ratio, $\equiv \omega_2 / \omega_1$

$\mu$ dynamics viscosity $Nm^{-2}\ s$

$\nu$ kinematic viscosity $m^2 s^{-1}$

$\rho$ density of fluid $kg\ m^{-3}$

$\tau$ shear stress $N\ m^{-2}$

$\phi$ dissipation rate $N\ m^{-2}\ s^{-1}$

$\omega_1$ angular velocity of the inner cylinder rad $s^{-1}$

$\omega_2$ angular velocity of the outer cylinder rad $s^{-1}$

$\omega_{1a}$ angular velocity of the inner cylinder after splitting rad $s^{-1}$

$\omega_{2a}$ angular velocity of the outer cylinder after splitting rad $s^{-1}$

## 1. Introduction

The flow device of Taylor-Couette Flow comprising concentric rotating cylinders is widely used in many industrial and research processes found in chemical, mechanical and nuclear engineering. The device can be only one cylinder rotating and the other at rest, or two cylinders rotating in the same or counter directions. The accurate calculation of the flow property is important even from the standpoint of the normal operation of the device. The distribution of energy loss in the device may greatly influence the industrial process of mixing, diffusion, heat transfer, and flow stabilities, etc. Despite the importance, it is interesting to note that the method for calculating or the actual calculation of the distribution of energy loss in such a device has not been found in the open literature. In particular, for the case of two cylinders rotating in the same direction, the calculation of energy loss distribution may pose some difficulties. In this note, the principle and the detailed derivation for the calculation are given.

As in many engineering problems, knowing about the loss distribution in the flow is very useful for enhancing the performance of the device and increasing the efficiency. For example, the design of an airfoil can be done according to the prescribed loss distribution or pressure distribution in order to increase the lift without compromising on the safety aspects. In the design of axial compressors, the deflection angle of the air flow passing a blade can be varied along spanwise direction in terms of the distribution of energy loss (along the height of the blade). In the design of centrifugal compressors, the distribution of energy losses is also used in the aerodynamic calculation and design of the three-dimensional impeller and the distorted vane diffusers for the purposes of enhancing the efficiency and of broadening the operation range [1-2]. Energy loss due to viscosity in the flow can reduces the efficiency of the fluid transportation. On the other hand, it may enhance the flow stability in some cases.

In the past years, the problem of flow between two concentric rotating cylinders has been extensively studied in terms of the flow stability due to infinitesimal disturbances [3-5]. This problem was first investigated experimentally by Couette and Mallock, respectively, in 1890s [3-5]. It was observed that the torque needed to rotate the outer cylinder increased linearly with the rotation speed until a critical rotation speed, after which the torque increased much more rapidly. This change was due to a transition from stable to unstable flow at the critical rotation speed. For the stability of an inviscid fluid moving in concentric layers, Rayleigh used the circulation variation versus the radius to explain the instability [6] and von Karman employed the relative roles of centrifugal force and pressure gradient to interpret the instability initiation [7]. Their goal was to determine the condition for which a perturbation resulting from an adverse gradient of angular momentum can lead to instability. In a classic paper, Taylor presented a mathematical stability analysis for viscous flow and compared the results to laboratory observations [8]. Taylor interpreted the experiment observation and linked to mathematical calculation for the instability initiation. Taylor observed that, for a gap between cylinders much smaller than the radii and under a given rotating speed of the outer cylinder, as the rotating speed of the inner cylinder exceeds a certain critical value, rows of cellular pattern are developed. In Taylor's findings, it is shown that the increase of fluid viscosity can delay the instability. These works have been considered as classical physics [9-10]. More detailed experimental study for the flow structure and pattern with the variations of the rotating speeds of the two cylinders have been given in literature [11-12].

Plane Couette flow is the limiting status of Taylor-Couette flow when the radii of the cylinders tend to towards infinite dimension. The former is linearly stable via eigenvalue analysis for all the Reynolds number, while the latter displays a critical value of the Taylor number from the classical linear stability analysis performed by Taylor [8]. How the linear instability that leads to the formation of Taylor vortices is lost in the case of the plane Couette flow is not known. Some authors try to associate or link these two flows using numerical simulation and experiments [13, 14]. Recent studies show that the instability in shear flows is dominated by the energy gradient in transverse direction and the energy loss in the streamwise direction for wall-bounded parallel flows [15, 16].

Generally, the energy loss due to viscosity in shear flows is helpful to delay the flow instability subjected to a perturbation such as found in Taylor-Couette flows and demonstrated by Taylor [8]. However, in the linear stability theory as found in literature [3,4,8], it is not shown where the flow instability is first started. Since the flow instability in shear flows is a local phenomenon and is intermittent in the beginning stage, it is interesting to find the position to first stimulate the flow instability. It is noticed that the analysis of energy loss in shear flows may provide some useful information for studying the flow instability. In pressure driven flows, the component of Laplace term in Navier-Stokes equation in streamwise direction represents rate of energy loss per volumetric fluid along the flow path. For plane Poiseuille flow, this rate of energy loss is constant along the transverse direction. For shear driven flows, this energy loss along the streamline is not explicitly shown in Navier-Stokes equations. In this paper, motivated from these ideas, distributions of the energy loss of unit volume of fluid along the streamline in the Taylor-Couette flows between concentric rotating cylinders are analyzed in detail for single cylinder rotation of inner or outer cylinder, and counter and same direction rotatiion of two cylinders. In order to present a link and comparison to plane Couette flow, the energy loss for plane Couette flow along the streamline is first derived for various flow conditions. The analytical results obtained can be helpful for understanding some complex phenomena occurring in the flow.

## 2. Mechanism of Energy Loss Enhancing Stability

In reference [15], Dou proposed a mechanism for flow instability and transition to turbulence in wall bounded shear flows. This mechanism suggests that the energy gradient in transverse direction plays a role of amplifying a disturbance and the energy loss in streamline direction serves the function of damping the disturbance. The related analysis obtains consistent agreement with the experimental data at the critical condition for wall bounded shear flows [15]. However, how the energy loss influences the stability has not yet demonstrated in detail. Below, we give an outline of this mechanism in principle.

In Fig.1(a), a fluid particle is located at the interface of two layers of fluid. The velocities for the two layers of fluid are $u$ and $u_B$ ($u < u_B$), respectively. Viscous stress is generated at the interface due to the momentum exchange of fluid particles between the two layers. From the equilibrium of forces, viscous stress is balanced by the inertial force and the pressure. From the conservation of energy, the energy loss generated by viscous friction is balanced by the energy drop (pressure driven flow) or the energy input (shear driven flow). The flow of the fluid particle in a steady laminar flow should be stable at finite Re unless it is subjected to disturbance(s).

In Fig.1(b), a fluid particle is located at lower layer with a velocity $u$ and kinetic energy $0.5\rho u^2$ . If this particle moves to the upper layer under a disturbance and then returns to its original streamline at the lower layer, this movement of the particle will cause its velocity and kinetic energy to change owing to the momentum and energy exchanges with other particles in the upper layer. This is because there is an energy gradient along the transverse direction. If there is no energy gradient in the transverse direction, this particle could not get energy from the movement. Meanwhile, this particle is subjected to energy loss along the flow path when this particle exchanges momentum with other particles during the movement (inelastic collision). We express its final velocity after the return to its original streamline as $u_1$ ($u < u_1 < u_B$), then this particle obtains a velocity increment $(u_1 - u)$ as compared to its original velocity. This variation forms a streamwise disturbance velocity $(u_1 - u)$ and a disturbance energy $0.5\rho(u_1^2 - u^2)$ of the base flow which is the genesis of the amplifying process of the disturbance development and can lead to flow transition when a threshold is achieved. The energy loss of this particle along the path during the movement reduces the magnitude of the velocity $u_1$ and thus reduces the magnitude of the disturbance energy obtained as $0.5\rho(u_1^2 - u^2)$ . This energy loss of the particle due to viscous friction should be proportional to the energy loss of the base flow in the vicinity of the original streamline. If the energy loss of base flow is larger, the value of the disturbance kinetic energy $0.5\rho(u_1^2 - u^2)$ obtained will become small, and vice versa. If the energy loss due to viscous friction is large enough (like the case of Re less than 2000 for pipe flow), all of the disturbance energy obtained during the cycle of disturbance will be damped out no matter how large the amplitude of the imposed disturbance. This is the reason why the energy loss of base flow enhances the flow stability.

It should be noticed from above discussion that the mechanism of flow instability in shear flows is due to the variation of kinetic energy of base flow resulting from the action of transversal disturbance. The velocity profile of the mean shear flow provides a background of leading to flow instability since it supplies an energy gradient in the transverse direction. The interaction of transversal disturbance with this energy gradient results in the change of kinetic energy of base flow which tries to destabilize the flow. On the other hand, the energy loss in the streamwise direction due to viscosity serves to damp out the outcome of the interaction, and it enhances the flow stability.

## 3. Energy Loss Distribution for Plane Couette Flow

### 3.1 One plate moving and the other plate fixed

In the plane Couette flow, the viscous term $\mu\nabla^2\mathbf{u}$ in the Navier-Stokes equations is zero, and the fluid energy $E = p + \frac{1}{2}\rho V^2$ per unit volume for incompressible fluid is constant along the streamwise direction. This is not to say that there is no energy loss due to viscous friction in the flow. The friction loss must still occur since this is a viscous fluid (the zero energy loss only occurs for inviscid flow). The energy level is kept constant because the energy loss due to friction is exactly compensated by the energy input to the flow by the moving wall. The work done to the flow by the moving wall is balanced by the energy loss in the flow. This can be obtained from the law of energy conservation. For a fluid tube along the streamline in incompressible flow, the law of energy conservation for per unit volume of fluid is

$$d(p + \frac{1}{2}\rho V^2 + \rho g z) = -dH + dW \ . \tag{1}$$

Here, H is the energy loss per unit volume of fluid along the streamline; W is the work input to per unit volume of fluid by external object/influence.

For the plane Couette flow, $dp = 0$, $d(\frac{1}{2}\rho V^2) = 0$ and $d(\rho g z) = 0$. Therefore, we obtain along the streamline

$$dH = dW \ .$$

Thus,

$$\frac{dH}{dx} = \frac{dW}{dx} \ . \tag{2}$$

This equation means that the energy loss per unit length is equal to the work done by external object.

The velocity distribution for the plane Couette flow can be obtained by solving the Navier-Stokes equation, as in [17]. Because of $v = 0$ and $\frac{\partial}{\partial x} = 0$, the said equation for steady flow reduces to

$$\frac{\partial}{\partial y}\left(\frac{\partial u}{\partial y}\right) = 0 . \tag{3}$$

For the case of the upper plate moving while the bottom plate is at rest (Fig.2a), the streamline velocity is obtained as

$$u = \frac{U}{h} y , \tag{4}$$

of which the velocity gradient is

$$\partial u / \partial y = U / h . \tag{5}$$

The shear stress is calculated as

$$\tau = \mu \partial u / \partial y = \mu U / h . \tag{6}$$

By taking an element in the fluid layer as shown in Fig.3, the work done to the element by the upper layer is

$$A_1 = \tau \cdot \Delta x \cdot \Delta z \cdot (u + \Delta u) dt ,$$

and the work done on the lower layer by the element is

$$A_2 = \tau \cdot \Delta x \cdot \Delta z \cdot u \, dt .$$

Therefore, the net work done on the fluid element is given as (noticing that there is no other energy input),

$$\Delta A = A_1 - A_2 = \tau \cdot \Delta x \cdot \Delta z \cdot \Delta u \, dt .$$

This quantity equals to the energy consumed by the fluid element, i.e., energy loss within this fluid layer. Then, the energy loss of the fluid element per unit length in the streamwise direction is

$$\frac{\Delta A}{\Delta x} = \tau \Delta z \Delta u dt .$$

Since the shear stress is uniform in the domain (Eq.(6)), the term in the above equation is uniform in the domain too for a fluid element. It is noticed that this term for a fluid element is also a Galilean Invariant because $\partial u / \partial y$ is Galilean Invariant. However, the fluid element near the upper plate has higher velocity and the fluid element near the bottom plate has lower velocity. Thus, the energy loss of fluid element per unit volume fluid may vary across the transverse direction. The fluid volume passing through *dy* depth in *dt* time is

$$\Delta Q = \Delta y \cdot \Delta z \cdot u\, dt\,.$$

Hence, the energy consumed by the fluid element per unit volume of fluid in the length of $\Delta x$ in streamwise direction (Fig.3) is

$$\Delta H = \frac{\Delta A}{\Delta Q} = \frac{\tau \Delta x \Delta z \Delta u dt}{\Delta y \Delta z u dt} = \frac{\tau \Delta u}{u \cdot \Delta y}\Delta x\,. \qquad (7)$$

The **energy loss per unit volumetric fluid in unit length along the streamwise direction** is given as

$$\frac{\Delta H}{\Delta x} = \frac{\tau}{u}\frac{\Delta u}{\Delta y}\,.$$

As the fluid element can be made infinitesimally small in a continuous sense, the **rate of energy loss along the streamline direction** is hence obtained from the above equation as

$$\frac{dH}{dx} \equiv \frac{\tau}{u}\frac{du}{dy}\,. \qquad (8)$$

It should be emphasized that this rate of energy loss per unit volumetric fluid in unit length along the streamwise direction is not a Galilean Invariant. It is calculated along the flow path in unit length and is independent of time. It is distinguished from the usual energy dissipation rate of unit volumetric fluid which is independent of the flow path and flow distance and is a Galilean invariant. For a plane Couette flow, the latter can be expressed as

$$\phi = \tau \frac{du}{dy}\,. \qquad (9)$$

Strictly, this term can be also obtained by dividing the energy consumed of a fixed fluid element in unit time with the ***fixed volume of the fluid element*** $\Delta V (= \Delta x \Delta y \Delta z)$ for plane Couette flow, i.e.,

$$\frac{\Delta A}{\Delta V \cdot dt} = \frac{\tau \Delta x \Delta z \Delta u dt}{\Delta x \Delta y \Delta z \cdot dt} = \frac{\tau \Delta u}{\Delta y} => \tau \frac{du}{dy} = \phi\,.$$

As is commonly known, the energy dissipation rate per unit volumetric fluid in unit time ($\phi$) is a Galilean invariant. Its relationship to Eq.(8) is given below. The energy dissipation rate per unit volumetric fluid in *dt* time is $\phi$ dt. This fixed fluid element in $dt$ time is translated a distance $u \cdot dt$. Therefore, the energy dissipation of the fluid element in unit length is just the energy loss in unit length (Eq.(8)) ,

$$\frac{\phi dt}{u \cdot dt} = \tau \frac{du}{dy} / u = \frac{\tau}{u} \frac{du}{dy}$$

The equivalence of Eq.(8) for a plane Poiseuille flow, for purpose of comparison, is given as (this equation can be obtained from the energy equation and momentum equation),

$$\frac{dH}{dx} \equiv -\frac{\partial \tau}{\partial y} = -\mu \frac{\partial^2 u}{\partial y^2}. \tag{10}$$

Therefore, for the pressure driven flows like a Poiseuille flow, the rate of energy loss per unit volumetric fluid along the streamline as given by Eq.(10) is indeed a Galilean Invariant. There are many fundamental differences between the pressure driven and shear driven flows.

For a given position in y direction (Fig.2a) in plane Couette flow, the energy loss per ***unit volumetric fluid*** along the streamline from $x_1$ to $x_2$ can be obtained via integration of Eq.(8) as

$$H = \int_{x_1}^{x_2} \left( \frac{dH}{dx} \right) dx = \int_{x_1}^{x_2} \left( \frac{\tau}{u} \frac{du}{dy} \right) dx. \tag{11}$$

This kind of formulation is familiar in the context of fluid dynamics for turbomachinery and the other power machines [1-2].

Introducing Eq.(4) to (6) into Eq.(8), we then have

$$\frac{dH}{dx} \equiv \frac{\tau}{u} \frac{du}{dy} = \frac{\mu}{u} \left( \frac{du}{dy} \right)^2 = \frac{\mu U}{yh} = \frac{\mu U}{h^2} \frac{h}{y} . \tag{12}$$

It can be seen from Eq.(12) that the magnitude of energy loss in unit volume is proportional to U and is inversely proportional to $h^2$. This equation is plotted in Fig.4 (also see Fig.2a for the flow geometry). It can also be seen that at the bottom wall, the energy loss of unit volume fluid is infinite for y=0. At the upper wall, the consumed energy of unit volume fluid along the streamwise direction is (y=h)

$$\frac{dH}{dx} = \frac{\mu U}{h^2}. \tag{13}$$

It can be observed from Fig.4 that the energy loss per unit volume fluid increases with decreasing y along the width of the channel and tends towards infinity on approaching the bottom wall. Because the energy loss has a damping role to any flow disturbance, the flow near the bottom wall is therefore strongly stable. Towards the top plate, the energy loss is lowest and the flow is therefore most possibly unstable.

In the plane Poiseuille flow, the energy loss of unit volume fluid is constant along the width of the channel; see Eq.(10). Thus, the damping role due to energy loss to the disturbance is the same along the channel width. This is one of the main differences between the plane Couette flow and the plane Poiseuille flow. This difference in behaviour or characteristic may play a role in the process leading to instability of the flow.

### 3. 2 Two plates moving in the opposite directions

In this case, the coordinates is shifted to the centerline of the channel if $U_1 = U_2$ (see Fig.2b). If $U_1 \neq U_2$, the coordinates should shift to the place where u=0. The velocity profile is

$$u = \frac{U_1 + U_2}{2h} y. \tag{14}$$

The velocity gradient is

$$\frac{\partial u}{\partial y} = \frac{U_1 + U_2}{2h}, \tag{15}$$

and the shear stress is

$$\tau = \mu \frac{\partial u}{\partial y} = \mu \frac{U_1 + U_2}{2h}. \tag{16}$$

By taking an element of fluid in the flow and using the same procedure as before, similar equation to Eq.(12) can be obtained,

$$\frac{dH}{dx} \equiv \frac{\tau}{u} \frac{du}{dy} = \frac{\mu}{u} \left( \frac{du}{dy} \right)^2 = \frac{\mu(U_1 + U_2)}{2yh} = \frac{\mu(U_1 + U_2)}{2h^2} \frac{h}{y}. \tag{17}$$

It can be seen from Eq.(17) that the magnitude of energy loss is proportional to $(U_1 + U_2)$ and is inversely proportional to $2h^2$. The distribution of energy loss is shown in Figs.5 and 6 for different combinations of $U_1$ and $U_2$. The energy loss increases with the decreasing magnitude of y along the width of the channel, and it tends to be infinite at the centreline. Thus, the flow at the centreline is strongly stable. Therefore, the turbulence initiation would be very unlikely to originate from the centreline locality; any small disturbance can be easily damped out due to the associated large energy loss. On comparing Fig.5 to Fig.6, it is found that by changing the magnitudes of velocities of the two plates, this highest energy loss position can be moved between the two plates. This phenomenon or observation can be very useful for some industrial purposes. The fluid particles near the walls always have the smallest energy loss. These locations are likely places where instability can occur first. Bottin et al's experiments have actually indicated that the instability of the flow first started at the moving wall [18].

From the results of the two cases above, it can be suggested that for plane Couette flow the position of extreme energy loss tending towards infinite is located at the point of zero velocity. However, it is not applicable for two plates moving in the same direction.

### 3.3 Two plates moving in the same direction

When the two plates are moving in the same direction (Fig.2c), the method for calculating the energy loss as shown for the above two cases by simply taking an element directly from the fluid can not be employed. This is because there is no null velocity in the flow. For this case, the flow can be decomposed into two simple flows, in which one has the total energy consumption as the original and the other has null energy consumption. When the frame decomposition is used, both the first and the second laws of thermodynamics should be followed (see appendix). For the first law of thermodynamics, the total energy consumed by the whole system is conserved no matter how the coordinates is selected for Galilean transformation. For the second law of thermodynamics, the direction of the energy transfer should not be changed after the frame splitting. In the original configuration (left picture in Fig.7), the top plate does the work on the fluid, and the fluid transfer the energy down layer by layer. The fluid in the lowest layer does the work on the bottom plate. Thus, the direction of energy transfer is from top to down. For the present problem, the flow has to be firstly decomposed into two simple parts (see Fig.7). The velocity profile is decomposed as a simple shear flow (maximum velocity $U_{1a} = U_1 - U_2$) and a uniform flow (velocity $U_b = U_2$). It can be seen that the direction of energy transfer after the splitting is not changed, .i.e., from top to bottom (middle picture in Fig.7). The total energy loss of the system equals to the sum from the two velocity profiles split. The energy loss for the simple shear flow (part a) can be calculated using the same method provided in section 2.1 for the case of only a single moving plate. The result is similar to equation (12),

$$\frac{dH}{dx} \equiv \frac{\tau_a}{u_a}\frac{du_a}{dy} = \frac{\mu}{u_a}\left(\frac{du_a}{dy}\right)^2 = \frac{\mu U_{1a}}{yh} = \frac{\mu U_{1a}}{h^2}\frac{h}{y}. \tag{18}$$

where $u_a = u - U_2$ and $U_{1a} = U_1 - U_2$.

For the part b in Fig.7, this is a uniform flow (rigid body moving) and the viscous friction is zero in the whole flow field. Thus, the energy loss is zero due to no viscous friction in the flow field. Therefore, the total energy loss for two plates moving in the same direction is the value as expressed by Eq.(18), which is the same as that for one plate moving and the other plate remained fix. The only issue is to change or set the correct magnitude of the velocity from $u$ to $u_a = u - U_2$. Strictly, this is equivalent to changing the coordinate system. That is, the frame of reference is moving with the flow in a uniform speed $U_2$. The energy loss in the new moving coordinate system is the same as that of the previous fixed coordinate system. The distribution of energy loss is shown in Fig.8. The energy loss increases with decreasing y, and it tends towards infinity at the bottom even if the velocity at the bottom plate is not zero. From the examples of plane Couette flow, it is found that the position of extreme energy loss occurs always at the location of lowest velocity.

### 4. Energy Loss Distribution for Taylor-Couette Flow

The solution of velocity distribution between concentric rotating cylinders can be found in many texts, e.g. [3-5]. Firstly, we define that the components of the velocity in tangential and radial directions are expressed as u and v, respectively. Assuming $v = 0$ and $\frac{\partial}{\partial \theta} = 0$, the Navier-Stokes equation in circumferential direction for steady flows reduces to

$$\frac{\partial}{\partial r}\left(\frac{\partial u}{\partial r} + \frac{u}{r}\right) = 0. \tag{19}$$

Integrating the above equation gives the solution of the velocity,

$$u = Ar + \frac{B}{r} \tag{20}$$

and

$$A=\omega_1\frac{(\eta^2-\lambda)}{\eta^2-1} \text{ and } B=\omega_1 R_1^{\,2}\frac{(1-\lambda)}{1-\eta^2}, \tag{21}$$

where $\eta = R_1 / R_2$ and $\lambda = \omega_2/\omega_1$. $R_1$ is the radius of the inner cylinder and $R_2$ is the radius of the outer cylinder. $\omega_1$ and $\omega_2$ are the angular velocities of the inner and outer cylinders, respectively.

### 4.1 Inner cylinder rotating and the outer cylinder fixed

#### 4.1.1 The energy loss due to friction

In Taylor-Couette flow, the viscous term $\mu\frac{\partial}{\partial r}\left(\frac{\partial u}{\partial r}+\frac{u}{r}\right)$ in Navier-Stokes equations is zero, and the energy $p+\frac{1}{2}\rho V^2$ is constant along the streamwise direction. The energy loss due to friction is exactly compensated by the energy input to the flow by the moving cylinder so that the energy level is kept constant. The work done on the flow by the cylinder is balanced by the energy loss in the flow.

The flow for inner cylinder rotating and the outer cylinder fixed is shown in Fig.9a. Taking an element in the fluid layer as shown in Fig.10, the work done on the element by the outer layer is

$$\begin{aligned} A_1 &= F_1(\Delta x+\Delta\Delta x)=(\tau+\Delta\tau)(\Delta s+\Delta\Delta s)\Delta z(u+\Delta u)dt \\ &=(\tau+\Delta\tau)(r\Delta\theta+\Delta r\Delta\theta)\Delta z(u+\Delta u)dt \end{aligned}.$$

The work done on the inner layer by the element is

$$A_2 = F_2\Delta x=\tau\cdot\Delta s\cdot\Delta z\cdot udt=\tau\cdot r\Delta\theta\cdot\Delta z\cdot udt.$$

Here, $\Delta s = r\Delta\theta$ is the length of the arc element, and $\Delta z$ is the depth in the axial direction. The net work done on the element is therefore

$$\begin{aligned} \Delta A &= A_1 - A_2 = (\tau+\Delta\tau)(r\Delta\theta+\Delta r\Delta\theta)\Delta z(u+\Delta u)dt \\ &-\tau\cdot r\Delta\theta\cdot\Delta z\cdot udt \\ &=\begin{pmatrix}\tau r\Delta\theta\Delta u+\tau\Delta r\Delta\theta u+\tau\Delta r\Delta\theta\Delta u+\Delta\tau r\Delta\theta u+\Delta\tau r\Delta\theta\Delta u\\ +\Delta\tau\Delta r\Delta\theta u+\Delta\tau\Delta r\Delta\theta\Delta u\end{pmatrix}\Delta z dt \end{aligned}.$$

The fluid volume passing through *dr* depth in *dt* time is

$$\Delta Q = \Delta r \cdot \Delta z \cdot u dt \,.$$

Thus, neglecting high order terms, the energy consumed by the element in unit volume fluid is hence

$$\Delta H = \frac{\Delta A}{\Delta Q} = \left( \frac{\tau}{u} \frac{\Delta u}{\Delta r} + \frac{\tau}{r} + \frac{\Delta \tau}{\Delta r} \right) r \Delta \theta \,. \tag{22}$$

The streamwise element length is $\Delta s = r\Delta\theta$. The gradient of the consumed energy (energy loss gradient) in streamwise direction is,

$$\frac{\Delta H}{\Delta s} = \frac{\tau}{u} \frac{\Delta u}{\Delta r} + \frac{\tau}{r} + \frac{\Delta \tau}{\Delta r} \,.$$

Thus, when $\Delta r$ tends to be infinitesimal

$$\frac{dH}{ds} \equiv \frac{\tau}{u} \frac{du}{dr} + \frac{\tau}{r} + \frac{d\tau}{dr} \,. \tag{23}$$

Since the equation

$$\frac{2\tau}{r} + \frac{d\tau}{dr} = 0 \,, \tag{24}$$

holds in cylindrical coordinates for Taylor-Couette flow, we have

$$\frac{dH}{ds} \equiv \frac{\tau}{u} \frac{du}{dr} - \frac{\tau}{r} \,. \tag{25}$$

This equation is equivalent to Eq.(26) below for a pressure driven Poiseuille flow between concentric cylinders (Dean flow),

$$\frac{dH}{ds} \equiv \frac{\partial \tau}{\partial r} + \frac{2\tau}{r} \,. \tag{26}$$

### 4.1.2 Distribution of energy loss in the flow

The velocity gradient can be obtained from Eq.(20),

$$\frac{\partial u}{\partial r} = A - \frac{B}{r^2} \,. \tag{27}$$

The shear stress is

$$\tau = \mu \left( \frac{\partial u}{\partial r} - \frac{u}{r} \right) = \mu \left[ \left( A - \frac{B}{r^2} \right) - \frac{1}{r} \left( Ar + \frac{B}{r} \right) \right] = -\mu \frac{2B}{r^2} \,. \tag{28}$$

Thus,

$$\frac{\tau}{r} = -\mu \frac{2B}{r^3}, \tag{29}$$

and

$$\frac{\tau}{u}\frac{du}{dr} = -\mu \frac{2B}{r^2}\left(Ar + \frac{B}{r}\right)^{-1}\left(A - \frac{B}{r^2}\right). \tag{30}$$

Introducing Eq.(29) and Eq.(30) into Eq.(25), the energy loss is

$$\frac{dH}{ds} \equiv \frac{\tau}{u}\frac{du}{dr} - \frac{\tau}{r} = -\mu \frac{2B}{r^2}\left(Ar + \frac{B}{r}\right)^{-1}\left(A - \frac{B}{r^2}\right) + \mu \frac{2B}{r^3}$$

$$= \mu \frac{2B}{r^2}\left[\frac{1}{r} - \left(Ar + \frac{B}{r}\right)^{-1}\left(A - \frac{B}{r^2}\right)\right] = \mu \frac{4B^2}{r^4}\left(Ar + \frac{B}{r}\right)^{-1}. \tag{31}$$

Further, introducing Eq.(20) and (21) into Eq.(31), then we have

$$\frac{dH}{ds} \equiv \frac{\tau}{u}\frac{du}{dr} - \frac{\tau}{r} = \mu \frac{4B^2}{r^4}\left(Ar + \frac{B}{r}\right)^{-1}$$

$$= \mu \frac{4}{r^4}\left[\omega_1 R_1^{\ 2} \frac{1-\lambda}{1-\eta^2}\right]^2 \left[\omega_1 \frac{\eta^2 - \lambda}{\eta^2 - 1} r + \frac{1}{r}\omega_1 R_1^{\ 2} \frac{1-\lambda}{1-\eta^2}\right]^{-1}$$

$$= 4\mu \frac{\omega_1 R_1}{r^2}\frac{R_1^{\ 2}}{r^2}\frac{(1-\lambda)^2}{(1-\eta^2)^2}\left[\frac{\eta^2-\lambda}{\eta^2-1}\frac{r}{R_1} + \frac{R_1}{r}\frac{1-\lambda}{1-\eta^2}\right]^{-1}$$

$$= 4\mu \frac{\omega_1 R_1}{h^2}\frac{R_1^{\ 4}}{r^4}\frac{(1-\eta)^2}{\eta^2}\frac{(1-\lambda)^2}{(1-\eta^2)^2}\left[\frac{\eta^2-\lambda}{\eta^2-1}\frac{r}{R_1} + \frac{R_1}{r}\frac{1-\lambda}{1-\eta^2}\right]^{-1}. \tag{32}$$

Equation (32) is used for calculating the energy loss. Although this equation is derived for the case of inner cylinder rotating while the outer cylinder is at rest, we will see in later sections that this equation is also suitable for the case of two cylinders rotating in counter directions.

At the inner cylinder ($r = R_1$), the energy consumed per unit volume fluid in unit length is

$$\frac{dH}{ds} \equiv \frac{\tau}{u}\frac{du}{dr} - \frac{\tau}{r} = 4\mu \frac{\omega_1 R_1}{h^2}\frac{(1-\eta)^2}{\eta^2}\frac{(1-\lambda)^2}{(1-\eta^2)^2}\left[\frac{\eta^2-\lambda}{\eta^2-1} + \frac{1-\lambda}{1-\eta^2}\right]^{-1}. \tag{33}$$

If the outer cylinder is at rest ($\omega_2 = 0$) and inner cylinder rotates ($\omega_1 \neq 0$), we have at the inner cylinder,

$$\frac{dH}{ds} \equiv \frac{\tau}{u}\frac{du}{dr} - \frac{\tau}{r} = 4\mu\frac{\omega_1 R_1}{h^2}\frac{(1-\eta)^2}{\eta^2}\frac{1}{(1-\eta^2)^2}$$

$$= 4\mu\frac{\omega_1 R_1}{h^2}\frac{(1-R_1/R_2)^2}{(R_1/R_2)^2}\frac{1}{(1-R_1^2/R_2^2)^2}$$

$$= 4\mu\frac{\omega_1 R_1}{h^2}\frac{(R_2-R_1)^2}{R_1^2}\frac{R_2^4}{(R_2-R_1)^2(R_2+R_1)^2}$$

$$= 4\mu\frac{\omega_1 R_1}{h^2}\frac{R_2^2}{R_1^2}\frac{R_2^2}{(R_2+R_1)^2}. \quad (34)$$

where $h = R_2 - R_1$ is the width between the cylinders.

When the ratio of the channel width between the cylinders to the radius of the inner cylinder, ($h/R_1$), tends to zero, we have

$$\frac{R_2^2}{R_1^2} \to 1 \text{ and } \frac{R_2^2}{(R_2+R_1)^2} = \frac{1}{4}. \quad (35)$$

Thus, Eq.(34) reduces to

$$\frac{dH}{ds} \equiv \frac{\tau}{u}\frac{du}{dr} - \frac{\tau}{r} = \mu\frac{\omega_1 R_1}{h^2} = \mu\frac{U_1}{h^2}. \quad (36)$$

This expression at the limit of infinite radii of cylinders is the same as that for plane Couette flow.

The distribution of energy loss calculated using Eq. (32) is depicted in Fig.11, for the case of inner cylinder rotating while the outer cylinder is at rest. In this figure, the radius ratio $\eta = 0.90$ is used. It can be seen that the energy loss increases with increasing r along the width of the channel, and it tends to be infinite at the surface of the outer cylinder. Thus, the flow at the outer cylinder is strongly stable. Therefore, any small disturbance in the locality is likely to be damped out. The fluid particles near the inner cylinder have the smallest energy loss. This becomes a possible locality where instability can first occur, as generally observed in experiments [3,5,8]. This behaviour has important implication for some industrial processes.

### 4.2 Inner cylinder fixed and the outer cylinder rotating

When the inner cylinder is fixed ($\omega_1 = 0$) and the outer cylinder is rotating ($\omega_2 \neq 0$), Eqs.(20) and (27) for the velocity distribution still hold. For this case, Eq.(21) can be rearranged as

$$A = \omega_2 \frac{1}{1-\eta^2} \quad \text{and} \quad B = \omega_2 R_2^{\ 2} \frac{\eta^2}{\eta^2 - 1}, \tag{37}$$

where $\eta = R_1 / R_2$, and $R_1$ is the radius of the inner cylinder and $R_2$ is the radius of the outer cylinder.

In this case, taking an element in the fluid flow (Fig.9b), and using the same procedure as before, the equation for calculating the energy loss can be derived. It is found that Eqs.(22) to (31) are still hold for this case. By substituting Eq.(20) and (37) into Eq.(31), then Eq.(38) is obtained,

$$\begin{aligned}
\frac{dH}{ds} &\equiv \frac{\tau}{u}\frac{du}{dr} - \frac{\tau}{r} = \mu \frac{4B^2}{r^4}\left(Ar + \frac{B}{r}\right)^{-1} \\
&= \mu \frac{4}{r^4}\left[\omega_2 R_2^{\ 2} \frac{\eta^2}{1-\eta^2}\right]^2 \left[\omega_2 \frac{1}{1-\eta^2} r + \frac{1}{r}\omega_2 R_2^{\ 2} \frac{\eta^2}{\eta^2-1}\right]^{-1} \\
&= 4\mu \frac{\omega_2 R_2}{r^2} \frac{R_2^{\ 2}}{r^2} \frac{\eta^4}{\left(1-\eta^2\right)^2} \left[\frac{1}{1-\eta^2}\frac{r}{R_2} + \frac{R_2}{r}\frac{\eta^2}{\eta^2-1}\right]^{-1} \\
&= 4\mu \frac{\omega_2 R_2}{h^2} \frac{R_2^{\ 4}}{r^4} \frac{(1-\eta)^2 \eta^4}{\left(1-\eta^2\right)^2} \left[\frac{1}{1-\eta^2}\frac{r}{R_2} + \frac{R_2}{r}\frac{\eta^2}{\eta^2-1}\right]^{-1}.
\end{aligned} \tag{38}$$

Equation (38) is used for calculating the energy loss for this case. At the outer cylinder ($r = R_2$), the energy consumed per unit volume fluid in unit length is,

$$\begin{aligned}
\frac{dH}{ds} &\equiv \frac{\tau}{u}\frac{du}{dr} - \frac{\tau}{r} = 4\mu \frac{\omega_2 R_2}{h^2} \frac{(1-\eta)^2 \eta^4}{\left(1-\eta^2\right)^2} \left[\frac{1}{1-\eta^2} + \frac{\eta^2}{\eta^2-1}\right]^{-1} \\
&= 4\mu \frac{\omega_2 R_2}{h^2} \frac{(1-\eta)^2 \eta^4}{\left(1-\eta^2\right)^2}.
\end{aligned} \tag{39}$$

Rewriting above equation, we have

$$\begin{aligned}
\frac{dH}{ds} &\equiv \frac{\tau}{u}\frac{du}{dr} - \frac{\tau}{r} = 4\mu \frac{\omega_2 R_2}{h^2} \frac{(1-\eta)^2 \eta^4}{\left(1-\eta^2\right)^2} \\
&= 4\mu \frac{\omega_2 R_2}{h^2} \frac{\left(1 - R_1 / R_2\right)^2}{\left(1 - R_1^{\ 2} / R_2^{\ 2}\right)^2} \left(R_1 / R_2\right)^4
\end{aligned}$$

$$= 4\mu \frac{\omega_2 R_2}{h^2} \frac{R_2{}^2}{(R_2 + R_1)^2} \left( \frac{R_1}{R_2} \right)^4 . \qquad (40)$$

where $h = R_2 - R_1$ is the width between the cylinders.

When the ratio of the channel width between the cylinders to the radius of the inner cylinder, ($h / R_1$), tends to zero, we have

$$\left( \frac{R_1{}^2}{R_2{}^2} \right)^4 \to 1 \text{ and } \frac{R_2{}^2}{(R_2 + R_1)^2} = \frac{1}{4} . \qquad (41)$$

Thus, Eq.(40) reduces to

$$\frac{dH}{ds} \equiv \frac{\tau}{u} \frac{du}{dr} - \frac{\tau}{r} = \mu \frac{\omega_2 R_2}{h^2} = \mu \frac{U_2}{h^2} . \qquad (42)$$

This expression at the limit of infinite radii of cylinders is also the same as that for plane Couette flow.

The distribution of energy loss calculated using Eq. (38) is depicted in Fig.12, for the case of outer cylinder rotating while the inner cylinder is at rest. It can be seen that the energy loss decreases with increasing r along the width of the channel, and it tends to be infinite at the surface of the inner cylinder. Thus, the flow at the inner cylinder is strongly stable, and the flow at the out cylinder is most unstable. The flow behaviour for this case is very different from that for the case of inner cylinder rotating and outer cylinder at rest. In this situation, Taylor cell vortices pattern is skipped/bypassed and the flow directly transits to turbulence like as in plane Couette flow when the critical condition is reached as found in experiments [5,19]. However, the flow in this situation has not received sufficient concern in the past as pointed by Donnelly [19].

### 4.3 Two cylinders rotating in counter directions

In this case, taking an element in the fluid flow (Fig.9c), and using the same procedure discussed in Section 4.1, an equation similar to Eq.(32) can be obtained,

$$\frac{dH}{ds} \equiv \frac{\tau}{u} \frac{du}{dr} - \frac{\tau}{r} = 4\mu \frac{\omega_1 R_1}{h^2} \frac{R_1{}^4}{r^4} \frac{(1-\eta)^2}{\eta^2} \frac{(1-\lambda)^2}{(1-\eta^2)^2} \left[ \frac{\eta^2 - \lambda}{\eta^2 - 1} \frac{r}{R_1} + \frac{R_1}{r} \frac{1-\lambda}{1-\eta^2} \right]^{-1}$$

(43).

The distribution of energy loss calculated with Eq.(43) is shown in Fig.13 and Fig.14 for different combination of $\omega_1$ and $\omega_2$. The energy loss decreases with decreasing radial position near the inner cylinder and increasing radial position near the outer cylinder; it tends to be infinite at one particular position between the cylinders. The location of this position depends on the ratio of the angular velocities of two cylinders. Therefore, by changing the angular velocities of the two cylinders, this position can be moved between the cylinders. Due to the infinite energy loss at this position, the flow at this point is strongly stable. Therefore, any small disturbances in the said locality are likely to be damped out. The fluid particles near the cylinder surfaces have the smallest energy loss. Therefore, the instability generally occurs on the cylinder surfaces. The occurrence of instability may take place first on the inner cylinder or the outer cylinder, depending on other factors such as influences from radius of cylinders and the magnitudes of the rotating speeds. If the flow at the inner cylinder exceeds the critical condition, Taylor vortex cell pattern may first occur along the inner cylinder. If the flow at the outer cylinder exceeds the critical condition, the flow near the outer cylinder may directly transit to turbulence. If both of the flow at the inner cylinder and the flow at the outer cylinder exceed their critical conditions, complex flow pattern may be formed.

### 4.4 Two cylinders rotating in same direction

When the two cylinders rotate in the same direction (Fig.9d), the method for calculating the energy loss used for above-mentioned two cases can not be directly employed. This is because there is no null velocity in the flow. As such, the energy loss can not be simply obtained as for the case of single cylinder rotating. In this case, the flow can be decomposed as two simple flows (see Fig.15). In doing the frame splitting, the first and the second laws of thermodynamics should be followed as similar to the case of the plane Couette flow with two plates moving in same direction (see Appendix). The velocity profile is decomposed into two parts (Fig.15): (a) the inner cylinder is rotating and the outer cylinder is at rest (part a); (b) rigid body rotating with $\omega_2$ (part b). Thus, the angular velocity in the flow field for part a is $\omega_a = \omega - \omega_2$. The angular velocities at the two cylinders for part a are $\omega_{1a} = \omega_1 - \omega_2$ and $\omega_{2a} = 0$, respectively at inner and outer cylinders. The total energy loss is the sum of energy losses of the two velocity profiles of rotating flows. Next, the energy loss for the part a can be calculated using the same method as that in section 4.1 for the case of only one cylinder rotating. The result obtained is similar to Eq.(32), and is given by

$$\frac{dH}{ds} \equiv \frac{\tau_a}{u_a}\frac{du_a}{dr} - \frac{\tau_a}{r} = 4\mu\frac{\omega_{1a}R_1}{h^2}\frac{R_1^{\,4}}{r^4}\frac{(1-\eta)^2}{\eta^2}\frac{(1-\lambda)^2}{\left(1-\eta^2\right)^2}\left[\frac{\eta^2-\lambda}{\eta^2-1}\frac{r}{R_1}+\frac{R_1}{r}\frac{1-\lambda}{1-\eta^2}\right]^{-1} \tag{44}$$

where $\lambda = \omega_{2a}/\omega_{1a}$, $\omega_{1a} = \omega_1 - \omega_2$, and $\omega_{2a} = 0$.

For the part b in Fig.15, this is a rigid body rotating flow and the shear stress is zero in the whole flow field. Therefore, the energy loss for the part b is zero due to no viscous friction. Thus, the total energy loss for two cylinder rotating in same direction can be calculated just by Eq.(44), which is the same as that for the inner cylinder rotating and the outer cylinder at rest (Eq.(32)). The only requirement is to change the magnitude of the angular velocity from $\omega$ to $\omega_a = \omega - \omega_2$. Strictly, this method is equivalent to the changing of the coordinate system. That is, the frame of reference is rotating with the flow in a uniform angular speed $\omega_2$. The energy loss in the new rotating coordinate system is the same as that in the old fixed coordinates.

In Fig.9d and Fig.15, it is assumed that the rotating speed of the inner cylinder is larger than that of the outer cylinder ($\omega_1 > \omega_2$). If the rotating speed of the inner cylinder is less than that of the outer cylinder ($\omega_1 < \omega_2$), similar method can be used.

The distribution of energy loss calculated using Eq. (44) is shown in Fig.16, for the case of two cylinders rotating in same direction. This picture is the same as Fig.11 except the normalized ordinate, i.e., $\mu\omega_1 R_1/h^2$ in Fig.11 being replaced by $\mu(\omega_1-\omega_2)R_1/h^2$ in Fig.16. Therefore, the behaviour of energy loss for the two cases is identical. It can be seen from Fig.16 that the energy loss increases with increasing r along the width of the channel, and it tends to be infinite at the surface of the outer cylinder. The flow at the inner cylinder has lowest energy loss. Thus, the flow at the outer cylinder is strongly stable and the flow near the inner cylinder is most unstable. From this case, it is found that even if the flow velocity is not zero at the outer cylinder, the energy loss also tends towards infinity. Therefore, summarizing all the studied four cases, it is found that there is always a location in Taylor-Couette flow at which the velocity is the lowest and the energy loss is towards infinity.

Taylor was able to determine the critical condition of instability in the flow between concentric rotating cylinders via a mathematical stability analysis for viscous flow [8]. Strictly, in Taylor's analysis, it has also included the influence of energy loss as indicated by the Taylor number although the energy loss distribution is not considered. However, the present theory does not in any way contradict Taylor's analysis. On other hand, the present analysis reveals a potential mechanism found in most flow problems. We surmise that the principle of loss distribution is universal for most if not all flow problems. We further suggest that the loss distribution plays a partial but important role in flow instabilities. It either strengthens or diminishes the likelihood of occurrence of flow instability in the flows according to the distribution.

## 5. Concluding Remarks

In this work, the method for calculating the energy loss distribution in the Taylor-Couette flow between concentric rotating cylinders has been proposed. The principle and the detailed derivation for the calculation are given for single cylinder rotation of either inner or outer cylinder, and counter and same direction rotation of two cylinders. The distribution of energy loss due to viscosity in plane Couette flow and Taylor-Couette flow between concentric rotating cylinders are derived and discussed for various flow conditions. The findings have potentially important bearings on the flow stability and turbulence transition and hence great significance in the relation to many aspects of processes like mixing and heat transfer and others. The findings will be helpful for clarifying some complex flow phenomena and useful for the design of related industrial devices.

For plane Couette flow, the flow on the surface of moving plate has lowest energy loss if only one plate is moving. The flow at this location has lowest damp mode in response to any disturbance imposed, and hence the possibility that instability may occur first. The position of the highest energy loss occurring at the location of *lowest velocity* implies the presence of strongest damping to any perturbation. Thus, the flow at this said position tends to be stable. By changing the speed of the two plates, this stable location can move between the two plates.

For Taylor-Couette flow between concentric rotating cylinders with one at rest, the flow on the surface of the rotating cylinder has lowest energy loss. The flow at the said position has therefore lowest damping mechanism in the response to any disturbance. The possibility exists that instability may occur first at this location. On the other hand, highest energy loss occurs at the location of lowest velocity in the flow. The corresponding presence of strongest damping mechanism at such position may imply the most stable region. If the inner cylinder is rotating and the outer cylinder is at rest, the flow at the inner cylinder is most unstable, and while the flow at the outer cylinder is most stable. If the inner cylinder is at rest and the outer cylinder is rotating, the flow near the outer cylinder is most unstable, and while the flow at the inner cylinder is most stable.

For the counter rotating cylinders, the position of largest energy loss is located between the cylinders. By changing the angular velocities of the two cylinders, this position can be shifted between the cylinders. The most unstable locations are at the rotating cylinders in terms of their speeds. The flow stability and the flow pattern depend on the geometry and the rotating speeds of cylinders relative to their critical conditions.

For two cylinders rotating in same direction, the behaviour of energy loss is similar to the case of only one cylinder rotating. When the energy loss is calculated, the velocity profile must be decomposed into two parts, of which one should be rigid body rotating. The total energy loss is the sum the losses of the two velocity profiles. It is found that even if there is no location of velocity being null, there is always a position at which the energy loss tends towards infinity.

By summarizing the results for plane Couette flow and Taylor-Couette flow, it is found in shear driven flows that there is always a point at which the velocity is the lowest and the energy loss is towards infinity. Owing to the strong damping role of energy loss to disturbance, the flow is most stable at the said location. This may be the reason for the stability of some type of vortex flows. On the other hand, the surface of a moving object or a rotating cylinder is the place where the flow is most unstable. Thus, this may suggest a most effective way of mixing in these areas. All these findings can be utilized in various industrial processes.

## Acknowledgements

The authors thank Dr. Xing Shi for helpful discussions. They also wish to thank the anonymous referees for their helpful comments.

## Appendix: Application of Frame Splitting

In this paper, the frame splitting is used for the cases of two plates moving in same direction in plane Couette flows and two cylinders rotating in same direction in Taylor-Couette flows. When the frame splitting is used, the selection of the coordinates is not arbitrary. The basic principle for the superposition is that the frame splitting must obey the first and second laws of thermodynamics.

The first law of thermodynamics states that energy is conserved; it can be neither created nor destroyed. For the application in this study, the first law of thermodynamics is the energy conservation law. The energy conservation can be written as follows for the problem considered,

$$W_t + Q = K_2 - K_1 + P_2 - P_1 . \qquad \text{(A1)}$$

Here, $W_t$ is the work done by external to the system; Q is heat added to the system by external; K is the kinetic energy; P is the potential energy. The subscript 1 and 2 express the state of the system. Following the first law of thermodynamics, when we do a frame decomposition, conservation of energy should be kept before and after the frame splitting. The value of total energy loss in the whole system is not altered by the frame splitting. When the flow is steady, this value is equal to the work done by all the external objects.

The second law of thermodynamics states that the energy can only be transferred from the region of high energy to that of low energy and the energy transfer is irreversible. Therefore, the direction of transfer of energy should not be changed if one carries out the frame splitting.

We must distinguish the "*energy loss along the streamwise direction (see Eq.(8) and Eq.(25))* " and the *"total work done on the whole system by all external objects."* The former is a local quantity and is coordinate dependent (not Galilean invariant), and the latter is a global quantity and is a Galilean invariant.

However, one may take note that the work done on a part of flow with open boundary or on a part of system (opened system) is not a Galilean invariant, which is dependent on the selected coordinates. The first law of thermodynamics is for a whole system or an enclosed control volume; it is not just for a part of opened system, and is not for an opened local region in the flow field. The first law of thermodynamics is a universal law, it is not dependent on the frame selected (Galilean frame). It means that the work done or energy input to the whole system by external forces is conserved, which is independent of the coordinates.

For the case of two cylinders rotating in the same direction, it shares some behavior as for the case of plane Couette flow moving in the same direction, from the view point of energy transfer. Thus, we can split the energy field into the two parts for these configurations, one with null energy consumption and one with the total energy loss. This splitting obeys both the first and the second laws of thermodynamics. Because we use a rotating coordinates in the splitting for the case of two cylinders rotating in the same direction in this study and the rotating coordinate is not a Galilean frame, we give the proof for the validation of such a frame transformation which obeys the first law of thermodynamics.

Now, let us apply the first law of thermodynamics to the Taylor-Couette flow for the case of the two concentric cylinders rotating in the same direction. The work done to the fluid by a rotating cylinder in unit time can be written as (for unit length in the axial direction)

$$W = M\omega = F\cdot r\cdot\omega = (\tau_w\cdot 2\pi r)\cdot r\cdot\omega = 2\pi\cdot r^2\cdot\omega\cdot\tau_w, \qquad \text{(A2)}$$

where $M$ is the torque exerted on the fluid by the rotating cylinder and $F$ is the friction force exerted on the fluid by the rotating cylinder. The shear stress is expressed as (Eq.(29) in our paper),

$$\tau = \mu\left(\frac{\partial u}{\partial r} - \frac{u}{r}\right) = \mu\left[\left(A - \frac{B}{r^2}\right) - \frac{1}{r}\left(Ar + \frac{B}{r}\right)\right] = -\mu\frac{2B}{r^2}, \qquad \text{(A3)}$$

where $B = \omega_1 R_1^{\,2}\dfrac{(1-\lambda)}{1-\eta^2} = \dfrac{R_1^{\,2}}{1-\eta^2}(\omega_1 - \omega_2)$, $\eta = R_1 / R_2$ and $\lambda = \omega_2/\omega_1$. $R_1$ is the radius of the inner cylinder and $R_2$ is the radius of the outer cylinder. $\omega_1$ and $\omega_2$ are the angular velocities of the inner and outer cylinders, respectively.

Thus, we have,

$$W = -2\pi\cdot r^2\cdot\omega\cdot\tau_w = -2\pi\cdot r^2\cdot\omega\cdot\mu\frac{2B}{r^2} = -2\pi\cdot\omega\cdot\mu\cdot 2B = C\cdot B\omega$$

$$= \frac{CR_1^{\,2}}{1-\eta^2}(\omega_1 - \omega_2)\cdot\omega = D(\omega_1 - \omega_2)\cdot\omega, \qquad \text{(A4)}$$

where $C = -4\pi\cdot\mu$ and $D = \dfrac{CR_1^{\,2}}{1-\eta^2}$ are two constant for given $R_1$ and $R_2$.

For the original configuration (left picture in Fig.15), the work input to the system by the inner cylinder (in unit length in the axial direction) is:

$$W_1 = D(\omega_1 - \omega_2)\cdot \omega_1 . \tag{A5}$$

The work done on the outer cylinder by the working fluid (in unit length in the axial direction) is:

$$W_2 = D(\omega_1 - \omega_2)\cdot \omega_2 . \tag{A6}$$

Thus, the energy consumed by the fluid in the system is

$$H_t = W_1 - W_2 = D(\omega_1 - \omega_2)\cdot(\omega_1 - \omega_2) = D(\omega_1 - \omega_2)^2 . \tag{A7}$$

For the splitting configuration (middle picture in Fig.15), The work input to the system by the inner cylinder (in unit length in the axial direction) is:

$$W_{1a} = D(\omega_1 - \omega_2)\cdot \omega_{1a} = D(\omega_1 - \omega_2)\cdot(\omega_1 - \omega_2) = D(\omega_1 - \omega_2)^2 , \tag{A8}$$

while the work done on the outer cylinder by the working fluid (in unit length in the axial direction) is:

$$W_{2a} = 0 . \tag{A9}$$

Thus, the energy consumed by the fluid in the system is

$$H_t = W_1 - W_2 = D(\omega_1 - \omega_2)^2 - 0 = D(\omega_1 - \omega_2)^2 . \tag{A10}$$

Therefore, comparing Eq.(A7) and Eq.(A10), it is found that the energy consumed by the fluid is the same before and after splitting. In other words, the (external) work on the fluid is conserved before and after splitting. As the result, the first law of thermodynamics is conserved after splitting.

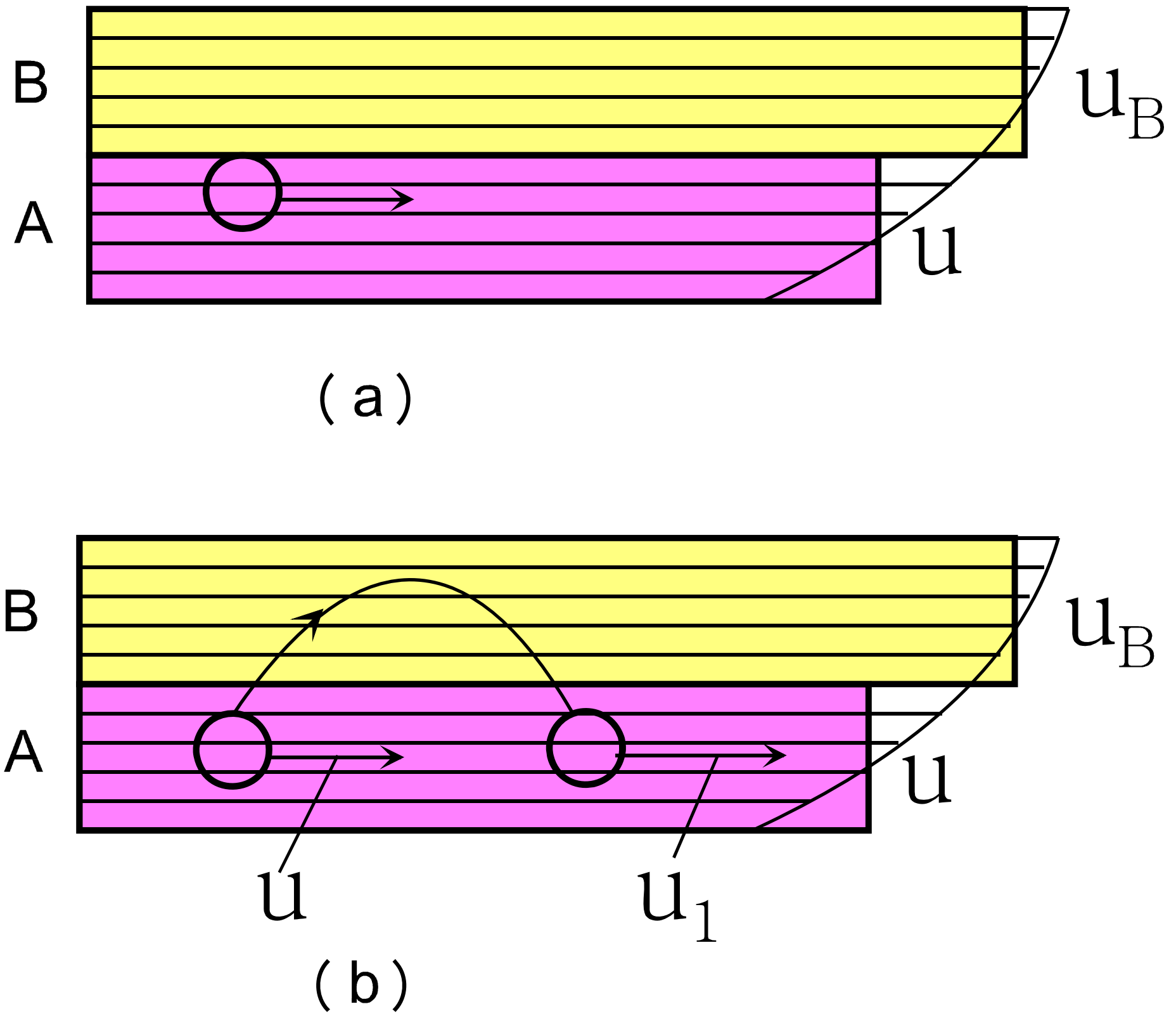

Fig.1 Skech for two layers of fluid in shear flows. (a) A particle flows stably in a fluid layer. (b) A particle undergoes a transversal disturbance. This particle obtains extra energy through the moving in the upper layer. Its energy at the end of moving is larger than its original energy due to the energy exchange at upper layer where the energy is high. This particle is also subjected to more energy loss (related to the energy loss of base flow) during the moving. The relative magnitude of the gained energy and the extra energy loss during the travelling in upper layer decides disturbance amplification and the stability of the flow.

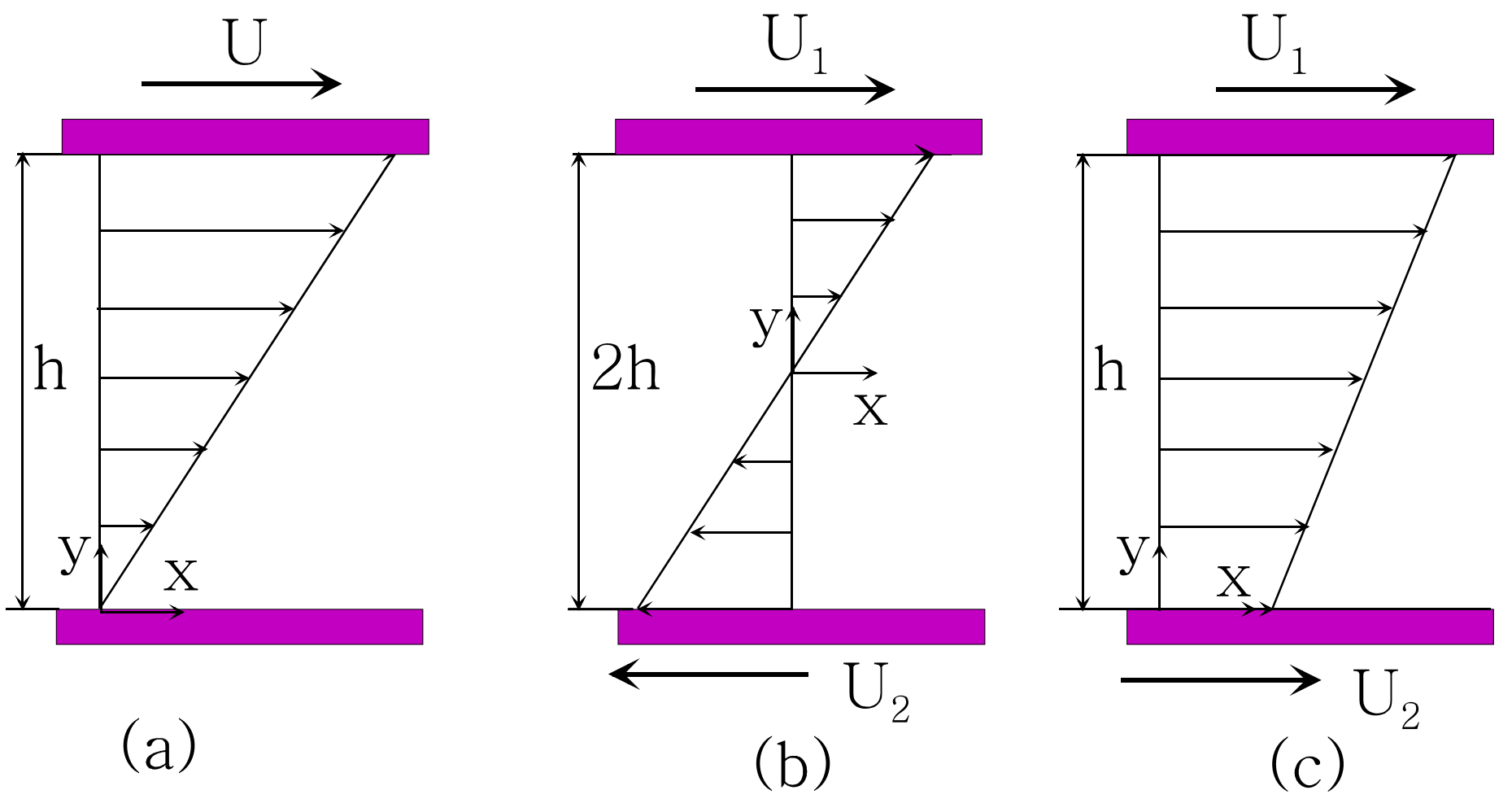

Fig.2 Plane Couette flow. (a) The bottom plate is at rest. (b) Two plates move in opposite directions. (c) Two plates move in same direction.

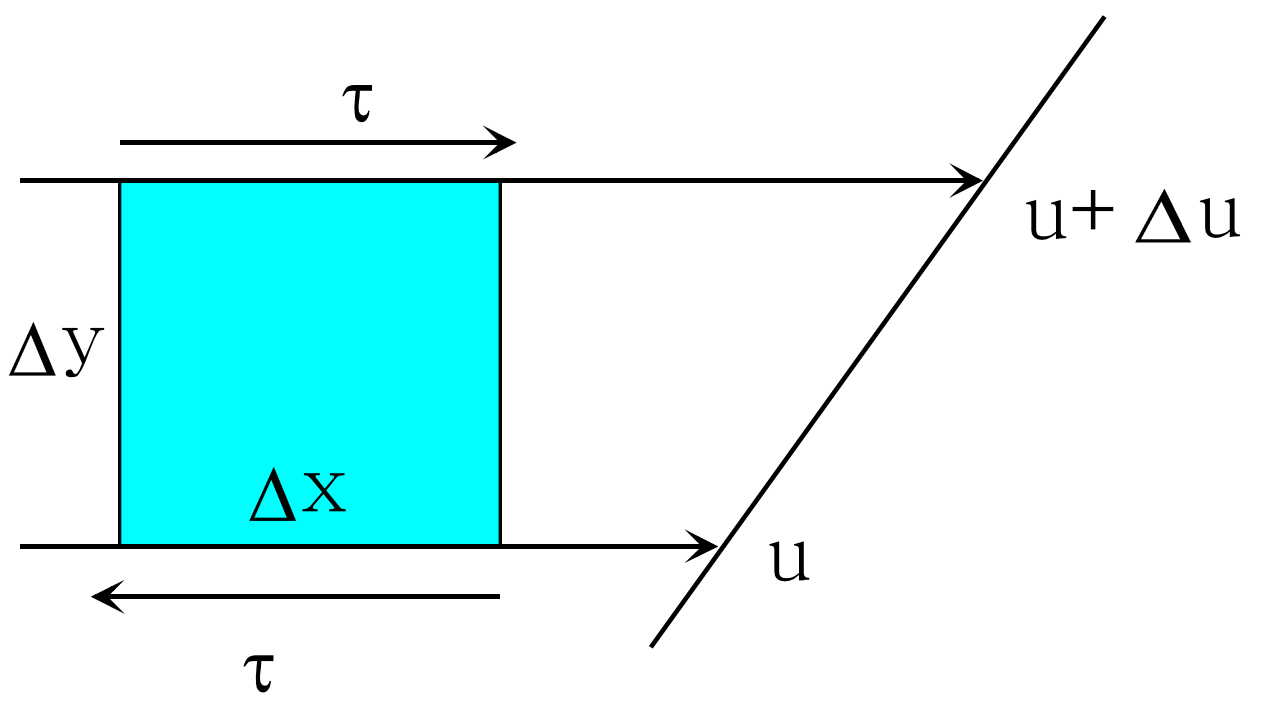

Fig.3 A cubic fluid element. Δz is perpendicular to x-y plane.

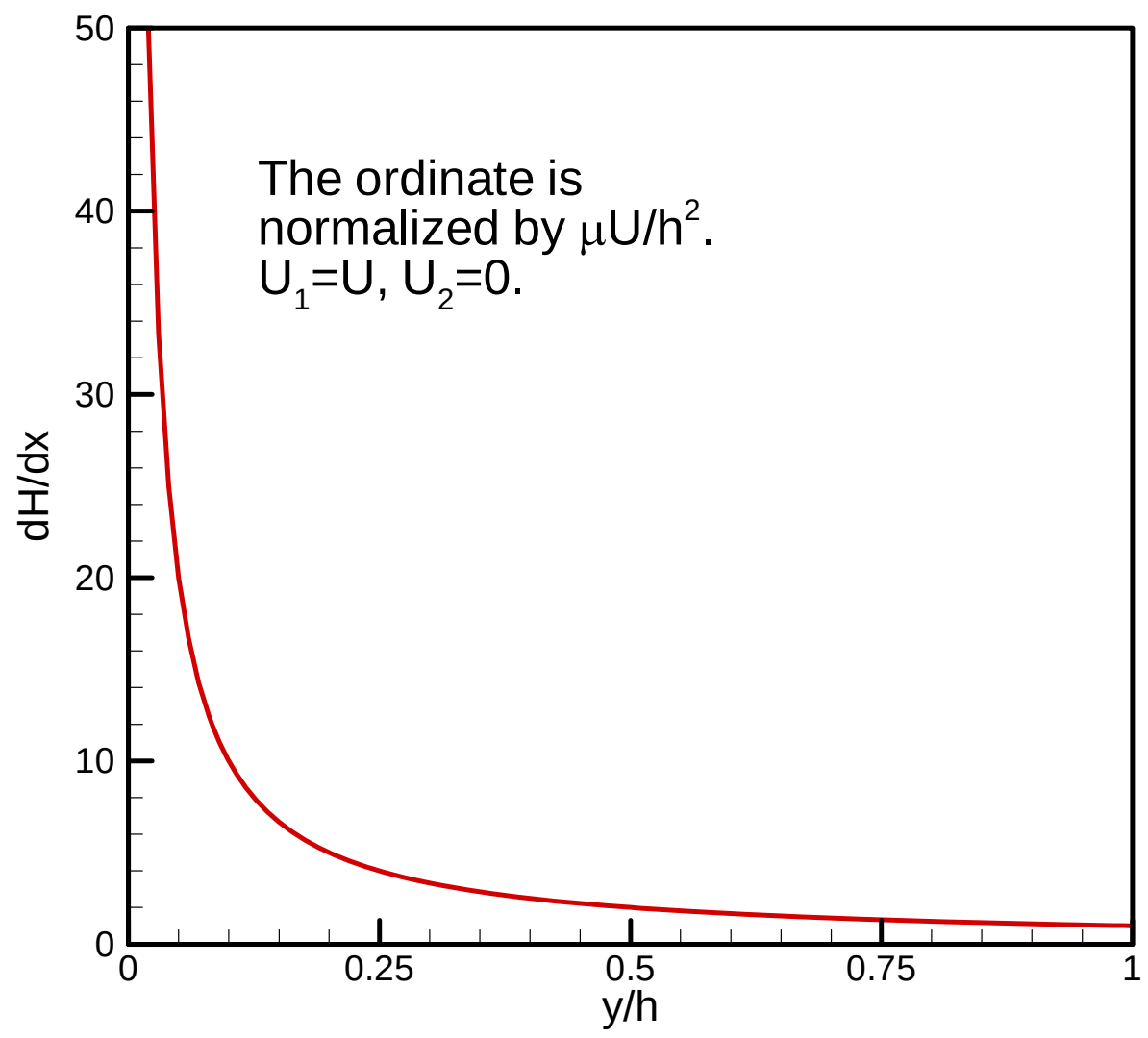

Fig.4 Energy loss along the channel width for plane Couette flow (only top plate moving).

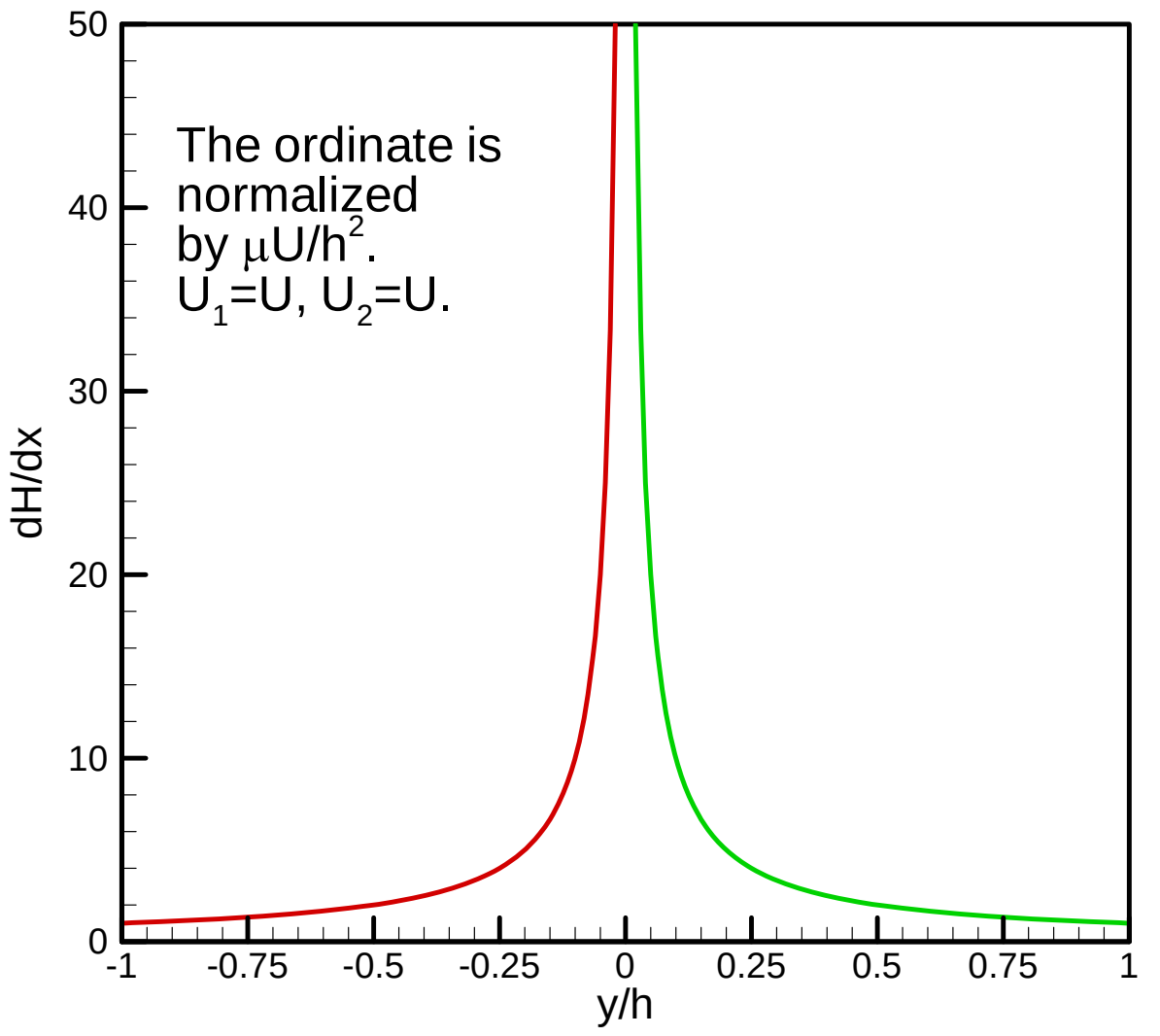

Fig.5 Energy loss along the channel width for plane Couette flow with two plates moving in opposite directions. The magnitudes of velocities of the two plates are the same.

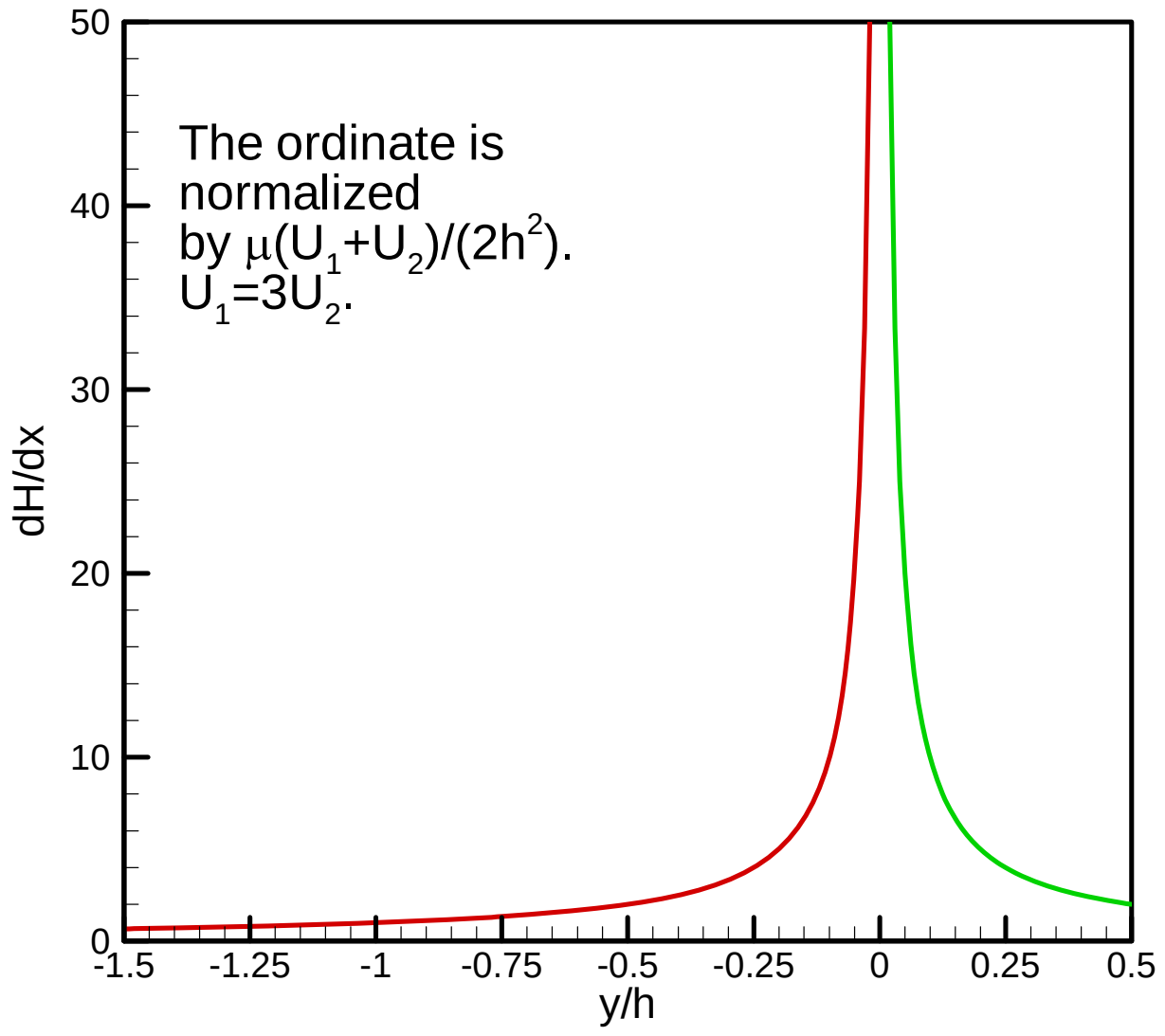

Fig.6 Energy loss along the channel width for plane Couette flow with two plates moving in opposite directions.

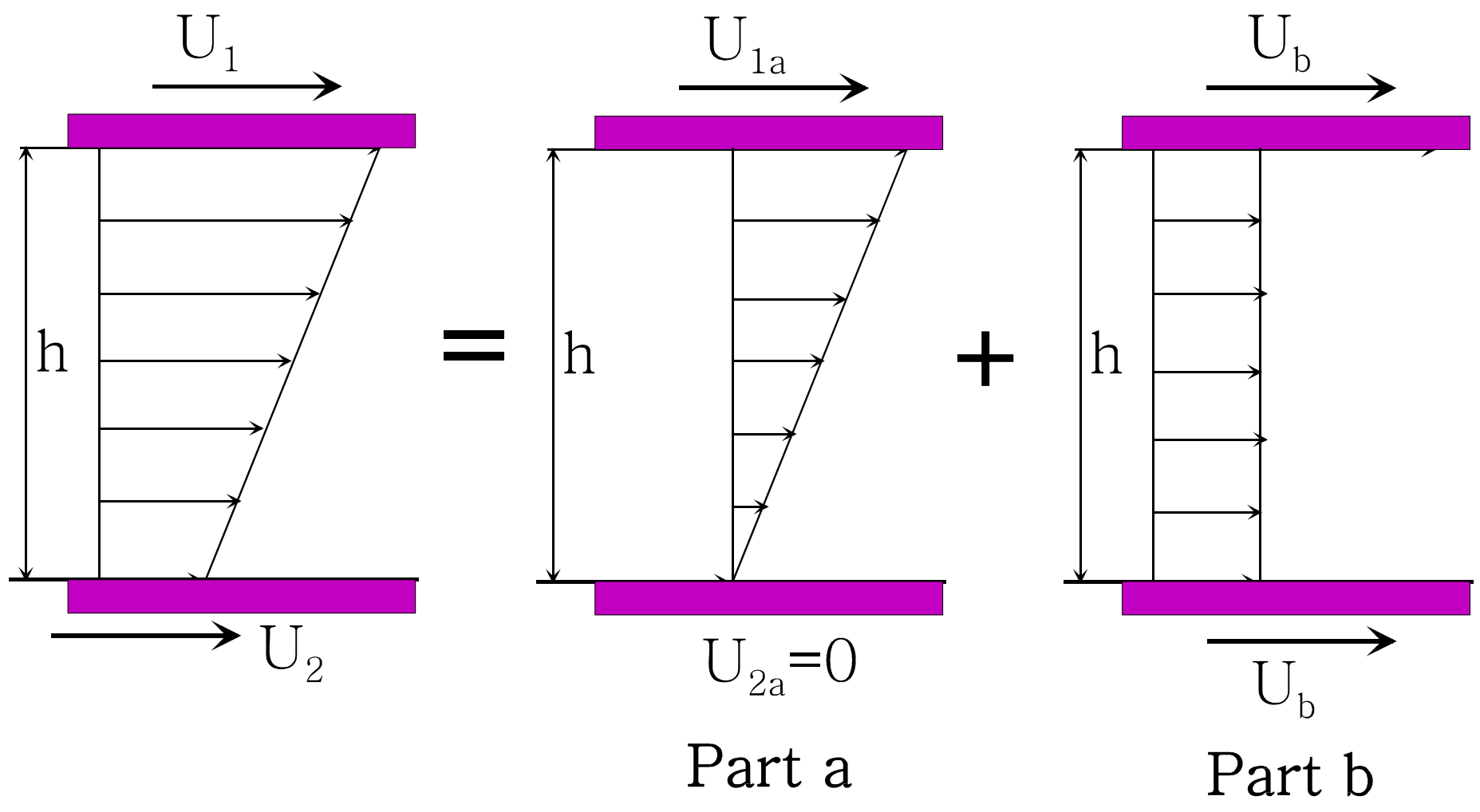

Fig.7 Velocity profile is decomposed into two profiles: Part a: simple shear flow; Part b: rigid body moving.

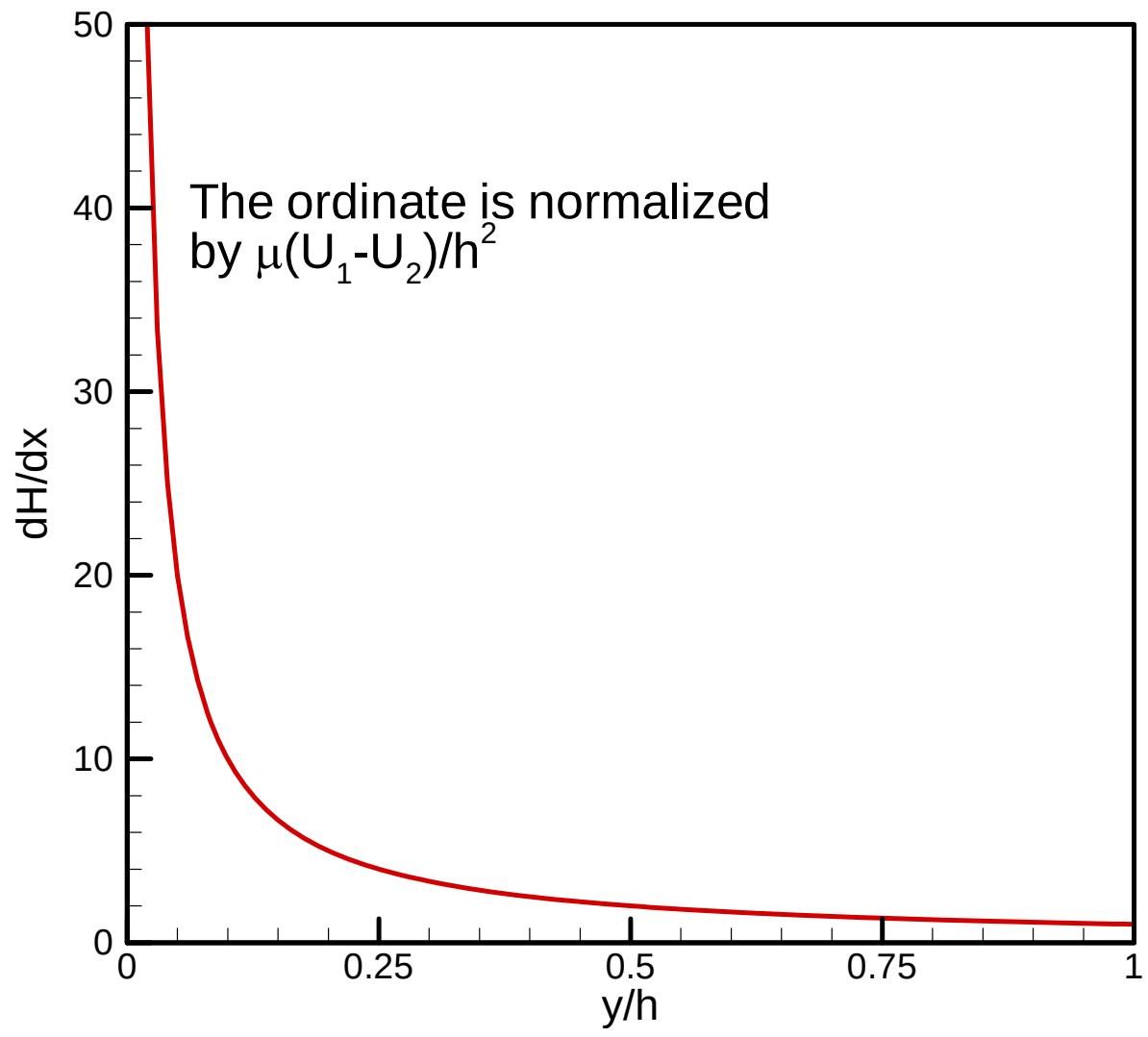

Fig.8 Energy loss along the channel width for plane Couette flow with two plates moving in same direction.

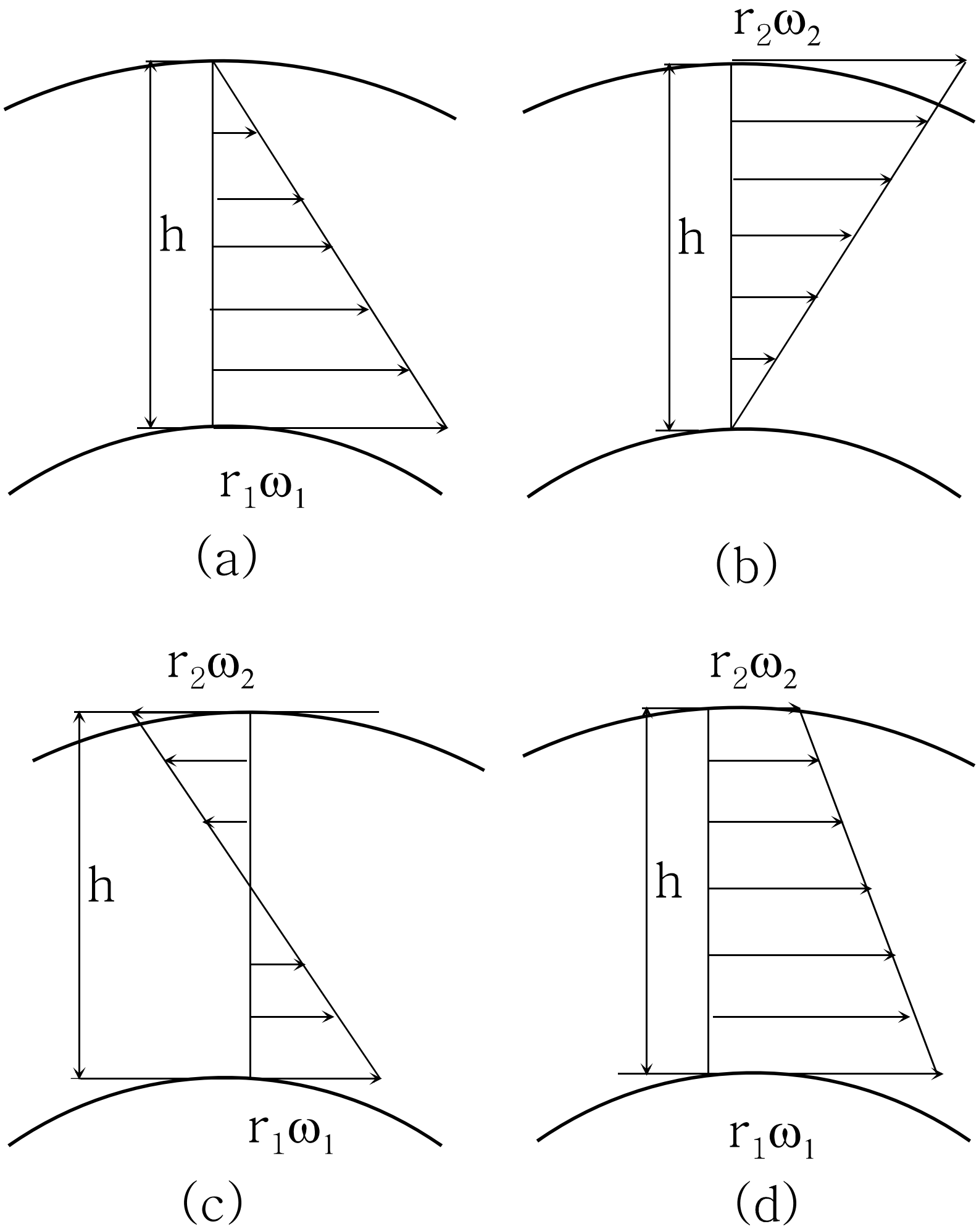

Fig.9 Velocity profile for concentric rotating cylinders; (a) Inner cylinder rotating and the outer cylinder at rest; (b) Inner cylinder at rest and the outer cylinder rotating; (c) Cylinders rotating in counter directions; (d) Cylinders rotating in same direction.

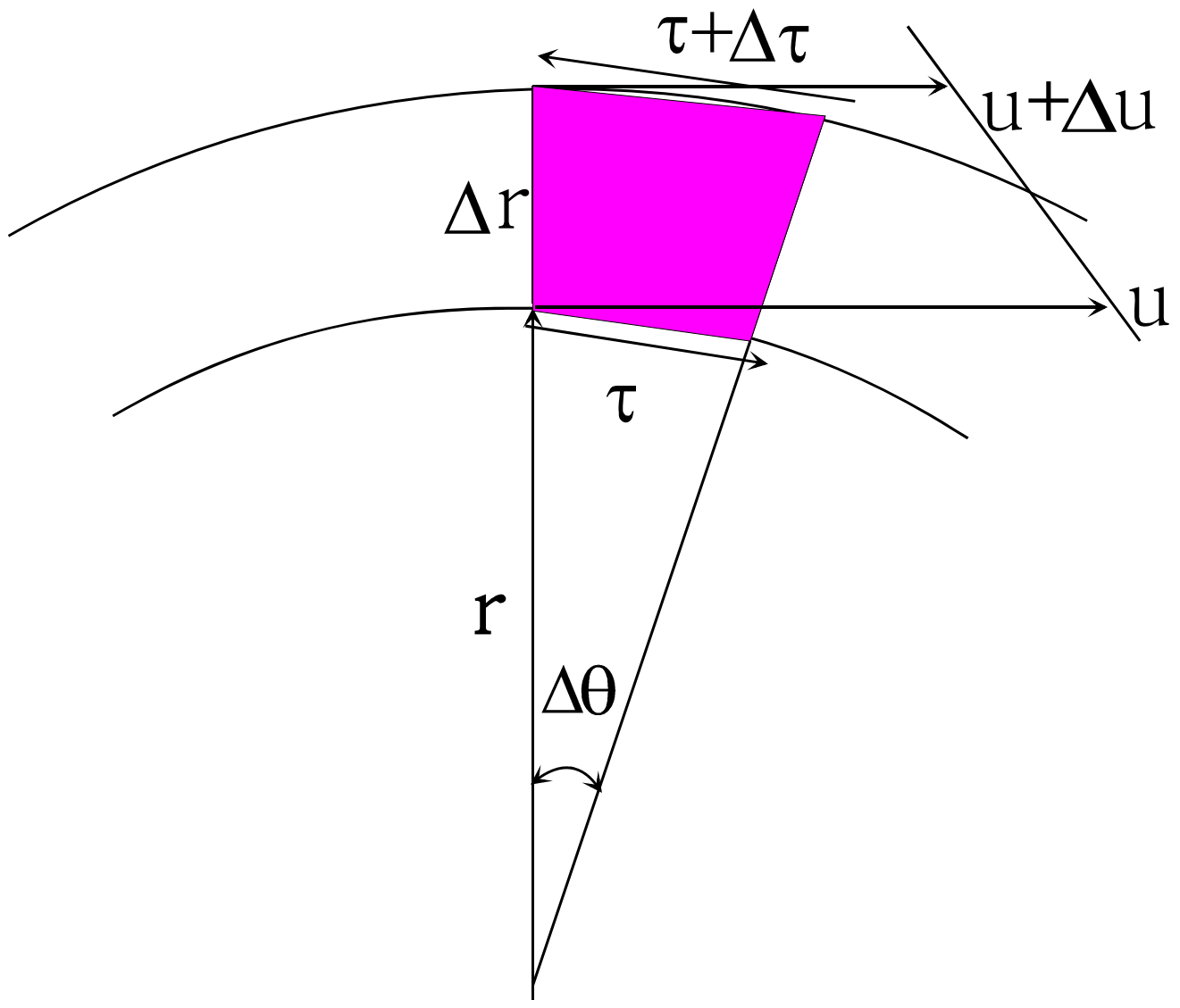

Fig.10 An annular fluid element taken from the flow between concentric rotating cylinders. $\Delta z$ is perpendicular to r-$\theta$ plane.

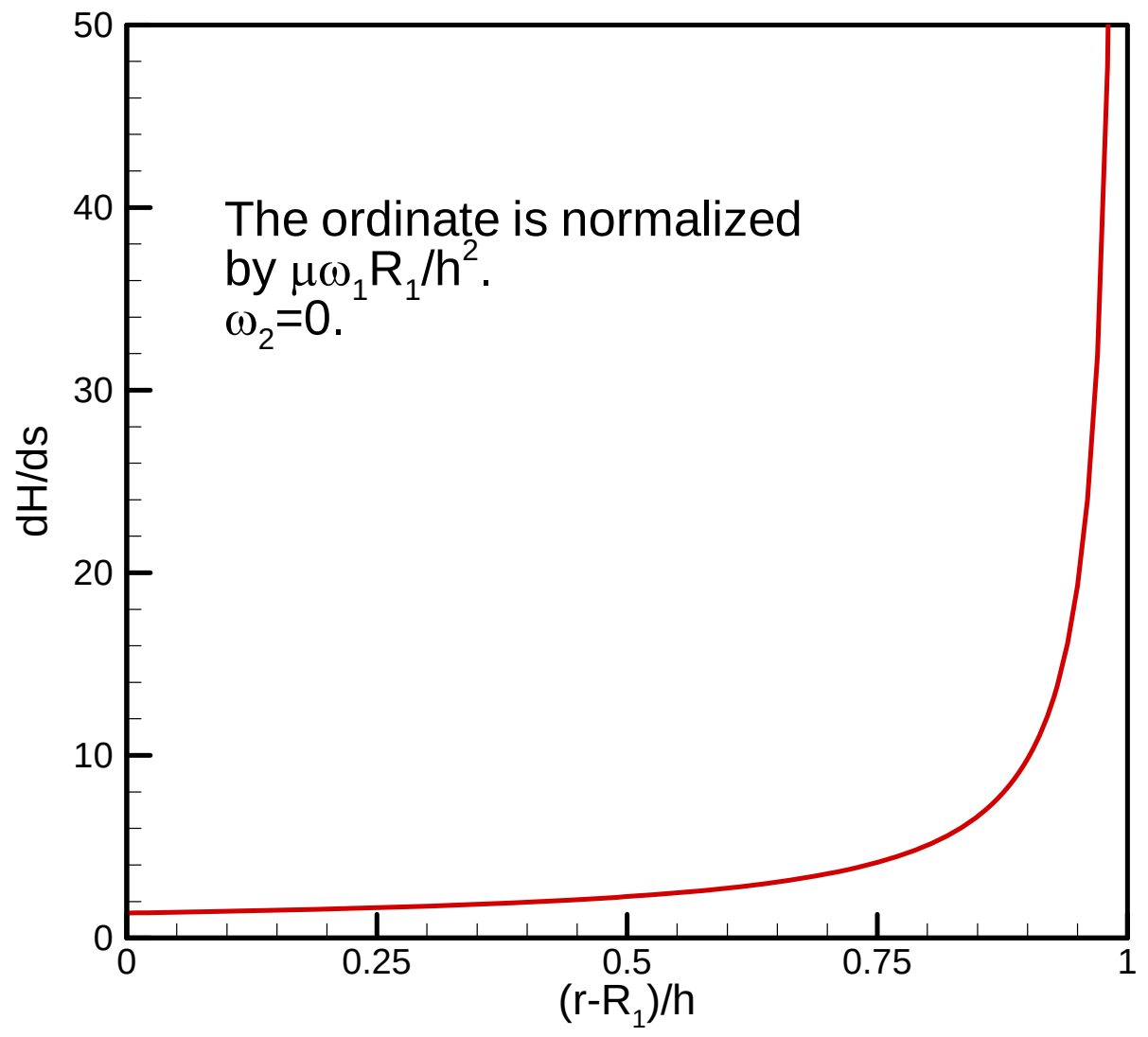

Fig.11 Energy loss along the channel width for concentric rotating cylinders (inner cylinder is rotating while outer cylinder is at rest). The radius ratio $\eta$=0.9 is used.

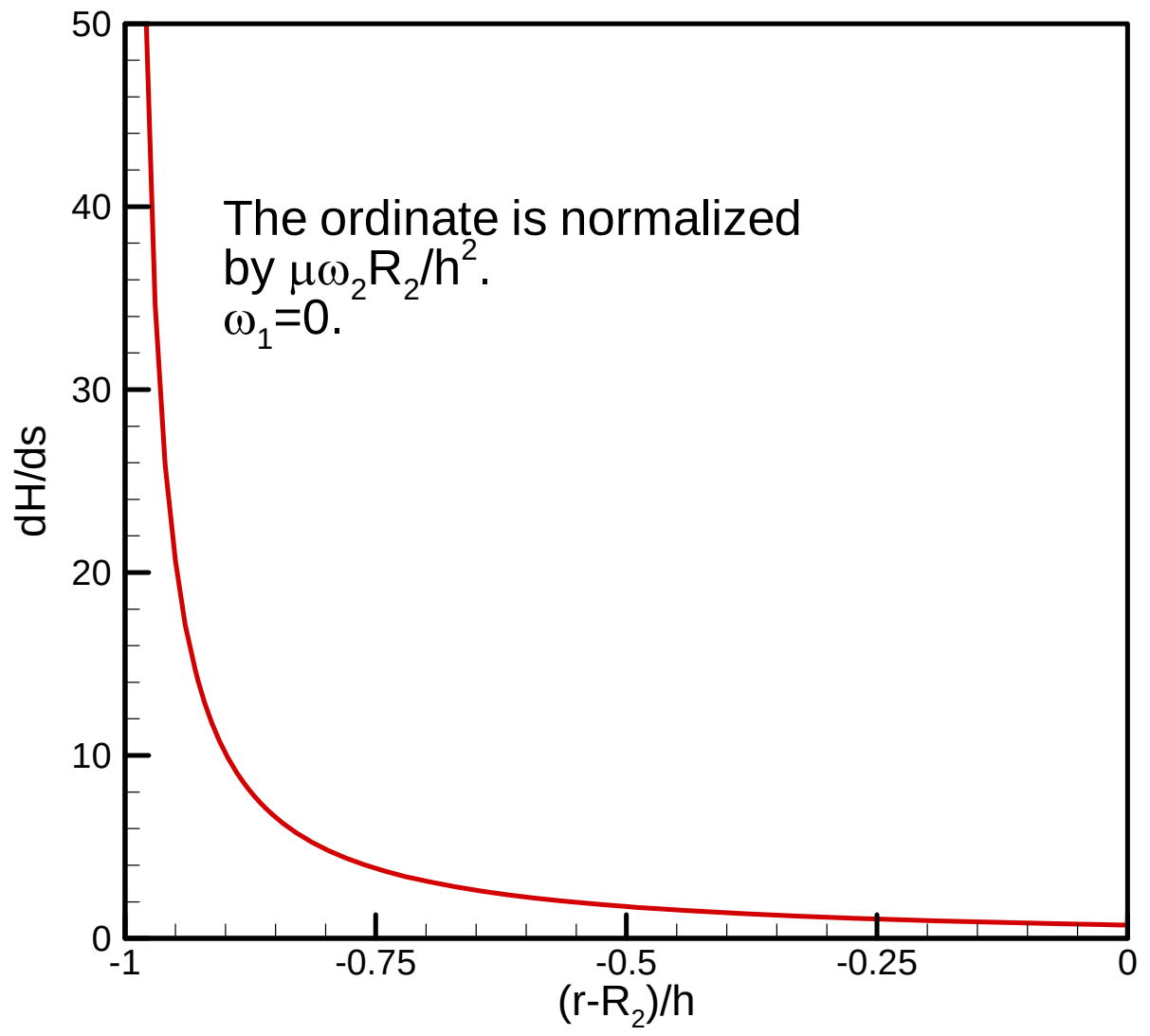

Fig.12 Energy loss along the channel width for concentric rotating cylinders (outer cylinder rotating while inner cylinder is at rest). The radius ratio $\eta$=0.9 is used.

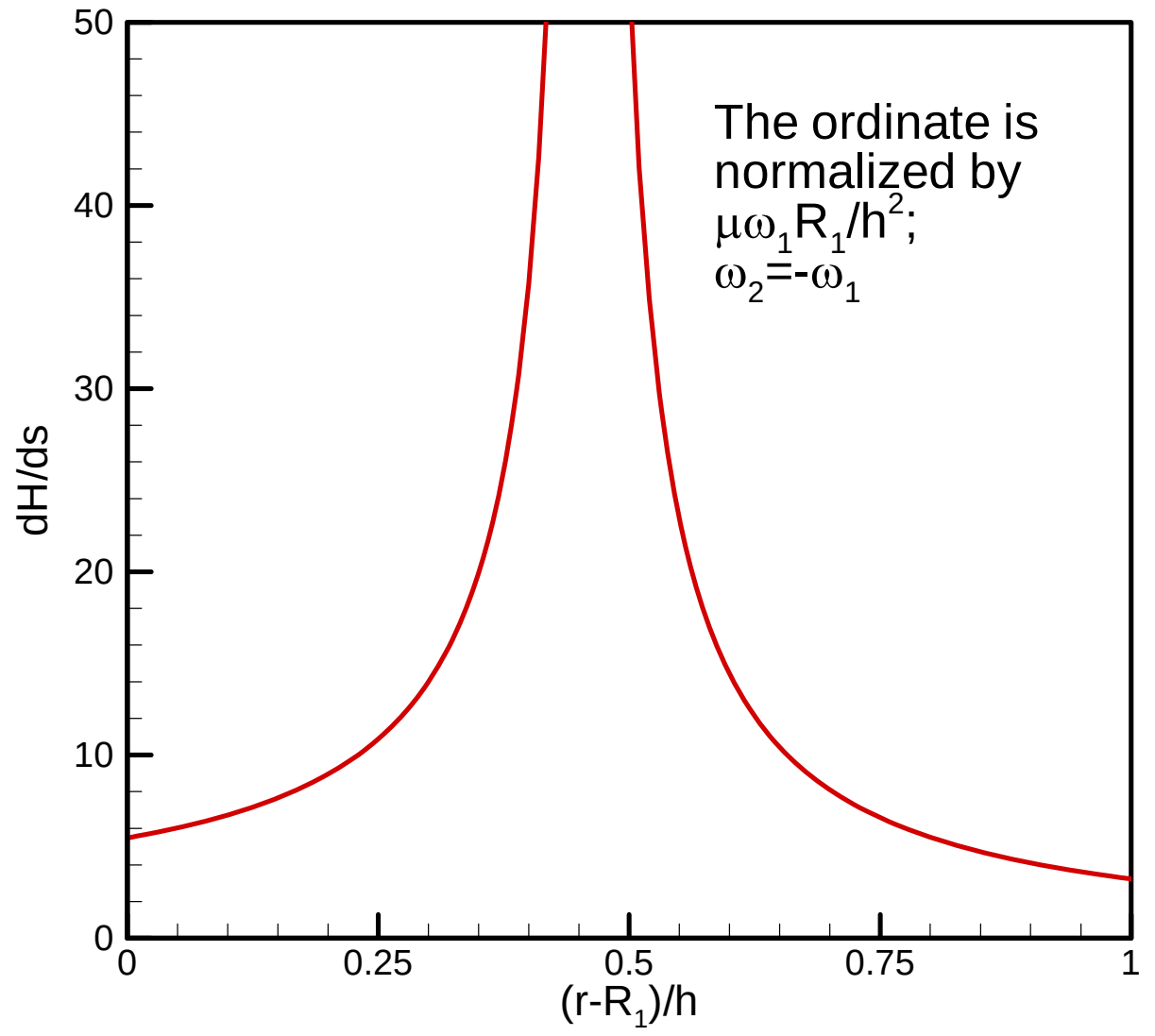

Fig.13 Energy loss along the channel width for concentric rotating cylinders with two cylinder rotating in opposite directions. The radius ratio η=0.9 is used.

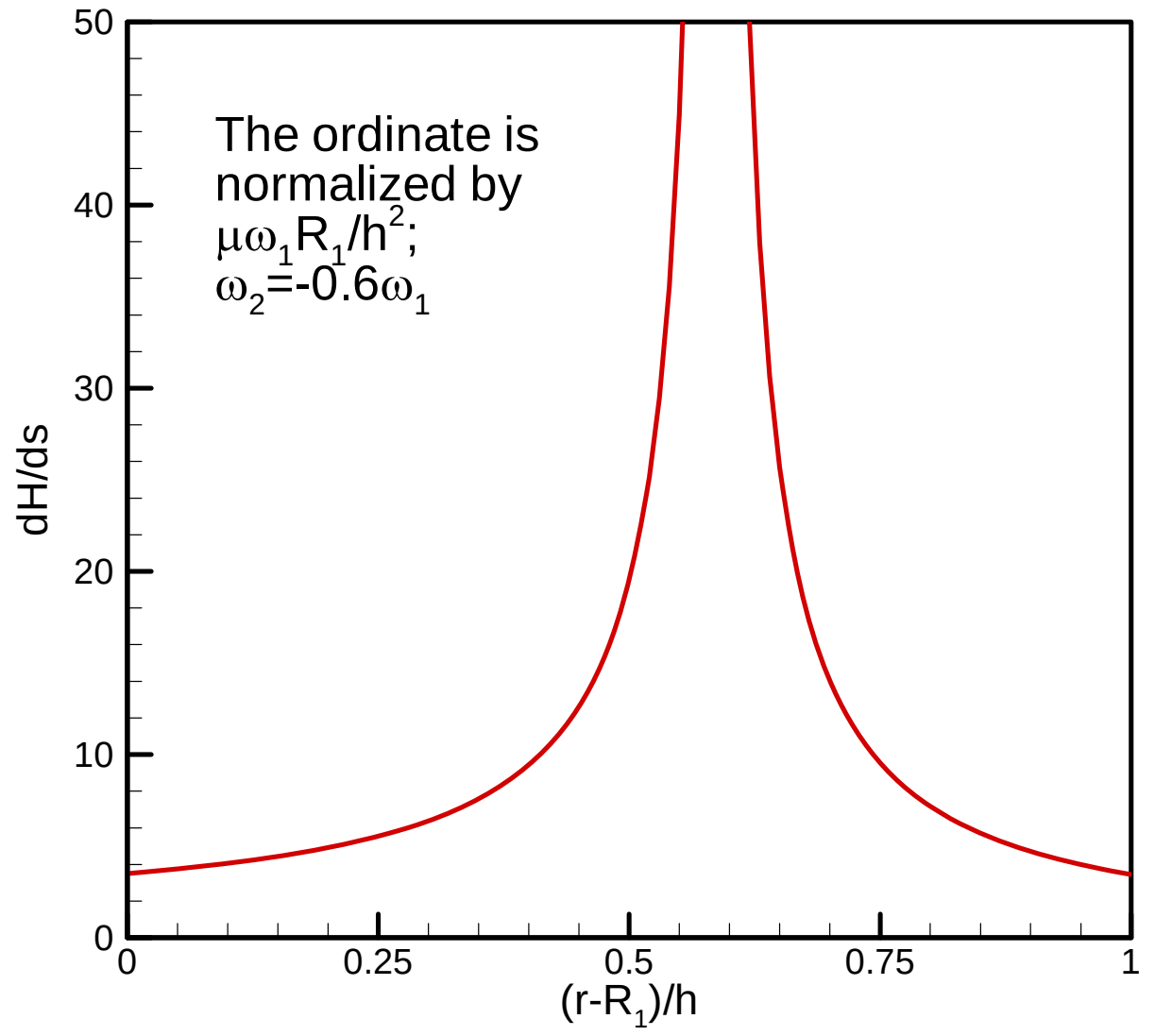

Fig. 14 Energy loss along the channel width for concentric rotating cylinders with two cylinders rotating in opposite directions. The radius ratio η=0.9 is used.

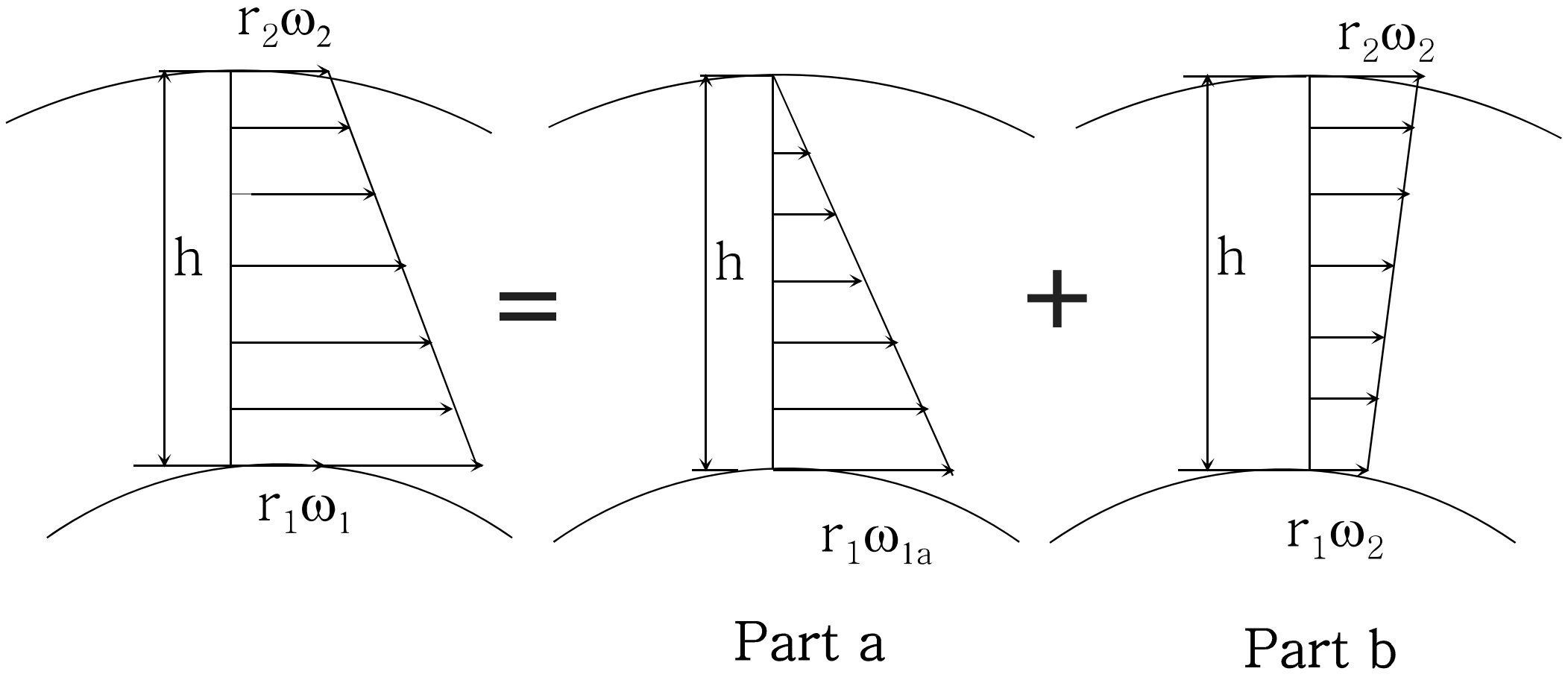

Fig.15 Velocity profile for cylinders rotating in same direction is decomposed into two profiles: Part a: outer cylinder at rest and inner cylinder rotating; Part b: rigid body rotating.

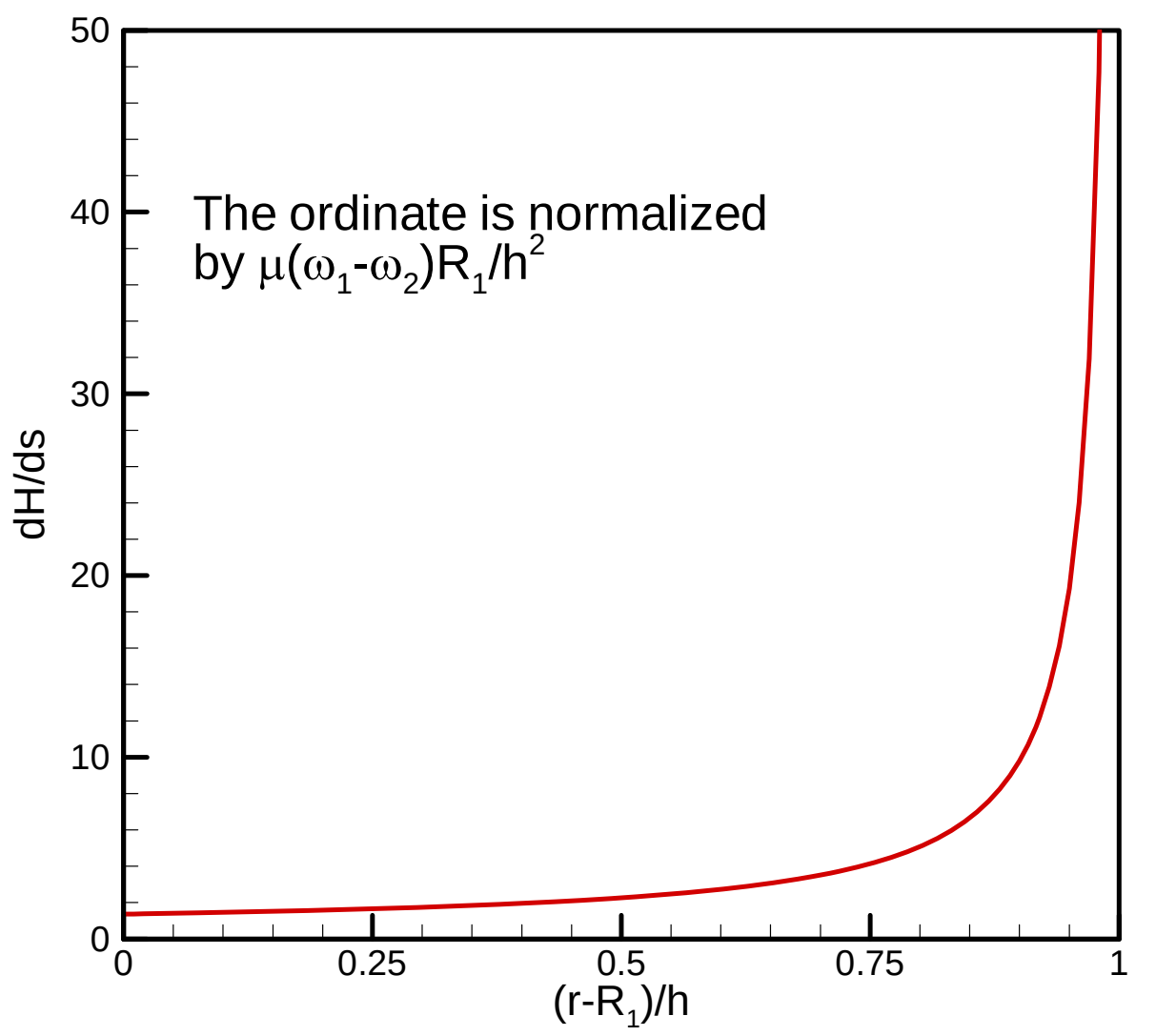

Fig.16 Energy loss along the channel width for concentric rotating cylinders with two cylinder rotating in same direction. The radius ratio $\eta$=0.9 is used.